%% file: template.tex
\documentclass[electronic]{vgtc}             

\graphicspath{{figures/}{pictures/}{images/}{./}} 
\usepackage{times}                     
\makeatletter
\def\thanks#1{\protected@xdef\@thanks{\@thanks
        \protect\footnotetext{#1}}}
\makeatother

\usepackage{tabu}                      
\usepackage{booktabs}                  
\usepackage{lipsum}                    
\usepackage{mwe}                       

\usepackage{mathptmx}                  
\usepackage{bbm}

\usepackage[T1]{fontenc}
\usepackage{amsmath}
\usepackage{amsfonts}
\usepackage{amssymb}
\usepackage{pifont}

\usepackage{makecell}
\usepackage{multirow}
\usepackage{multicol}
\usepackage{multibib}
\usepackage{booktabs}
\usepackage{duckuments}
\usepackage[export]{adjustbox}
\usepackage{subfigure}
\usepackage{bbding}
\usepackage{xcolor}

\usepackage{cite}  

\usepackage{mathptmx}                  

\newcommand{\xmark}{\ding{55}}

\onlineid{3984}

\vgtccategory{Technique}

\vgtcinsertpkg

\preprinttext{To appear in an IEEE VGTC sponsored conference.}

\ieeedoi{xx.xxxx/TVCG.201x.xxxxxxx}

\title{NARVis: Neural Accelerated Rendering for Real-Time Scientific Point Cloud Visualization}

\author{
        Srinidhi~Hegde,
        Kaur~Kullman,
        Thomas~Grubb,
        Leslie~Lait,
        Stephen~Guimond,
        and~Matthias~Zwicker
\thanks{
\\
\textbullet\ S. Hegde, M. Zwicker are with the University of Maryland, College Park.\\
\textbullet\ K. Kullman is with the University of Maryland, Baltimore County.\\
\textbullet\ T. Grubb, L. Lait are with NASA.\\
\textbullet\ S. Guimond is with Hampton University.\\
\textbullet\ Correspondence author e-mail: srihegde@umd.edu
}
}
\teaser{
  \centering
  \includegraphics[width=\linewidth]{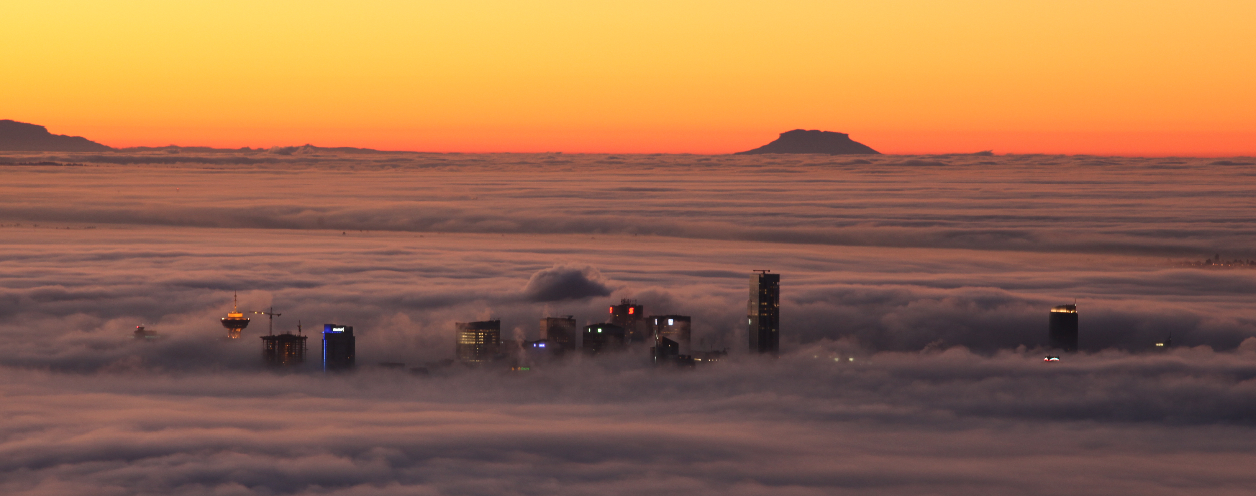}
  \caption{In the Clouds: Vancouver from Cypress Mountain.}
  \label{fig:teaser}
}

\input{sections/abstract.tex}

\teaser{
 \includegraphics[width=\linewidth]{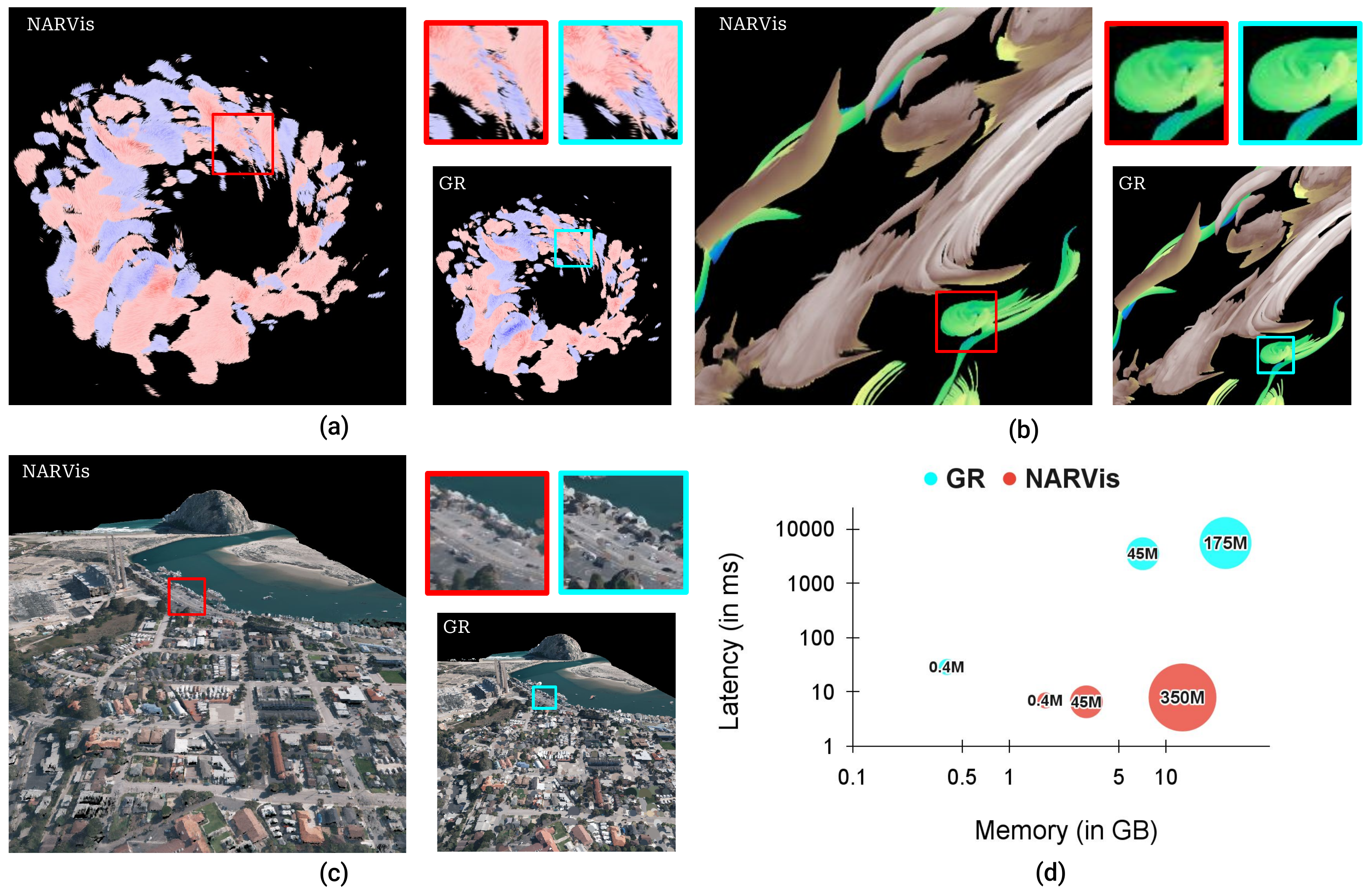}
 \centering
  \caption{We present a neural accelerated renderer, NARVis, for high-quality visualization of large scientific point clouds in real-time. NARVis can visualize a variety of scientific use cases, such as (a) static vector fields (\textit{Hurricane}), (b) particle trajectories (\textit{Storms}), and (c) photometric terrain scans (\textit{Morro Bay}). (d) NARVis renders high-quality images comparable to the conventional high-quality renderers (such as a Gaussian Renderer (GR), see section \ref{sec:train_data} for details on the renderer) with significantly lower latency and memory footprints. (Note: the axes are in log-scale and the bubble sizes denote the labeled number of points.)}
 \label{fig:teaser}
}

\graphicspath{{figs/}{figures/}{pictures/}{images/}{./}} 

\begin{document}



\maketitle

\input{sections/abstract.tex}
\input{sections/introduction.tex}
\input{sections/relworks.tex}
\input{sections/method.tex}
\input{sections/experiments.tex}
\input{sections/conclusions.tex}

\acknowledgments{%
	The authors would like to thank Roger Eastman for his generous support in providing the high-performance infrastructure necessary for this work. This work was funded by the National Aeronautics and Space Administration.%
}
\bibliographystyle{abbrv-doi-hyperref}
\bibliography{template}

\clearpage
\newpage
\appendix
\input{sections/appendix.tex}

\end{document}

%% file: sections/abstract.tex
\abstract{
Exploring scientific datasets with billions of samples in real-time visualization presents a challenge - balancing high-fidelity rendering with speed. This work introduces a neural accelerated renderer, NARVis, that uses the neural deferred rendering framework to visualize large-scale scientific point cloud data. NARVis augments a real-time point cloud rendering pipeline with high-quality neural post-processing, making the approach ideal for interactive visualization at scale. Specifically, we render the multi-attribute point cloud using a high-performance multi-attribute rasterizer and train a neural renderer to capture the desired post-processing effects from a conventional high-quality renderer. NARVis is effective in visualizing complex multidimensional Lagrangian flow fields and photometric scans of a large terrain as compared to the state-of-the-art high-quality renderers. Extensive evaluations demonstrate that NARVis prioritizes speed and scalability while retaining high visual fidelity. We achieve competitive frame rates of $>$ 126 fps for interactive rendering of $>$ 350M points (i.e., an effective throughput of $>$ 44 billion points per second) using $\sim$12 GB of memory on RTX 2080 Ti GPU. Furthermore, NARVis is generalizable across different point clouds with similar visualization needs and the desired post-processing effects could be obtained with substantial high quality even at lower resolutions of the original point cloud, further reducing the memory requirements.

}

%% file: sections/introduction.tex
\section{Introduction}



Real-time visualization is integral to the scientific community, offering immediate insights and enhancing decision-making for researchers dealing with complex datasets. Accurate data reproduction, precise rendering, responsive interaction, and coherent visual structures are essential for effective visualizations. However, conventional rendering techniques often fail to maintain the necessary throughput when handling massive point clouds, especially from high-throughput simulators~\cite{geos, maksimova2021abacussummit}, leading to high latency that hinders the investigative process. While this performance gap is most acutely felt in immersive display systems~\cite{simsick}, used in extended reality based visualizations~\cite{keefe2008scientific,millais2018exploring}, it remains a critical barrier for standard interactive workstations.
\\



\noindent Direct point cloud (PC) rendering methods and tools~\cite{SCHUETZ-2020-MPC,SCHUETZ-2021-PCC,CCsoftware} offer high accuracy within the limits of screen resolution and machine precision. However, these methods often lack the flexibility to incorporate diverse visual constructs, such as glyphs and textures, which is crucial in enhancing the data interpretability~\cite{interrante2000harnessing}. For example, the illustrative visualization paradigm~\cite{bujack2020state,brambilla2012illustrative} focuses on effective strategies for visualizing complex flow phenomena, like Lagrangian flow fields in fluid simulations (see Figure \ref{fig:teaser} (a), (b)), and requires rendering the velocity streaks, streamlines, transparency aware textures, etc. To this end, tools like QGIS~\cite{QGIS_software} and VAPOR~\cite{li2019vapor} offer various rendering presets to accommodate different visual styles but fail to render styles outside the preset domains and also struggle to scale to massive scientific datasets.
\\




\noindent Neural rendering is another category of rendering methods~\cite{aurand2022efficient,berger2018generative} that learns to adaptively and efficiently render stylized visualizations from diverse data representations. Neural rendering methods find applications in a variety of tasks in computer graphics such as interactive scene editing~\cite{aliev2020neural}, novel view synthesis, and animation synthesis~\cite{thies2019deferred}. Furthermore, a few works~\cite{edelstein2019, xiao2020neural} demonstrate the ability of neural rendering in the high-fidelity reconstruction of frames by reducing aliasing and accelerating the rendering frame rates in interactive video game applications.
\\

\noindent Inspired by the success of neural rendering methods for high-quality and real-time applications, in this work, we propose an efficient neural rendering framework, called \textit{NARVis}, that learns to modify the forward-rendered PC visualizations to include the required visual constructs for high-quality scientific data visualization. We initially render with a compute shader-based multi-attribute PC rasterizer, which we call Multi-Attribute Compute Rasterizer (MACR), and further refine with a neural network. Our rasterizer handles the workload of multi-variate data of high-resolution PC data from massive scientific simulations. Furthermore, we demonstrate that our method can visualize a variety of scientific PCs with various desired visualization styles, such as static vector fields, particle trajectories, and visual photometric scans. In this work, our primary contributions are as follows:

\begin{itemize}
    \item \textbf{Efficient Scientific Data Rendering:} We introduce an efficient multi-attribute PC rasterizer for accurately managing multiple data attributes from vast scientific PC datasets.
    \item \textbf{High-Quality Neural Post-Processing:} NARVis seamlessly blends various visual effects into the visualizations with end-to-end learnable neural rendering constructs.
    \item \textbf{Extensive Evaluation:} We illustrate the advantages of our framework over state-of-the-art methods to improve performance and rendering quality through extensive evaluation on various PCs.
\end{itemize}

%% file: sections/relworks.tex
\section{Related Works}

\subsection{Scientific Point Cloud Visualization}
Augmenting PC data with diverse attributes helps to visualize and understand various phenomena in different fields. Augmenting meteorological parameters (temperature, wind speed, humidity, etc.) for atmospheric vector fields  visualization~\cite{djurcilov2000visualizing,rautenhaus2017visualization}, adding kinematic properties of subatomic particles in collider experiments in physics~\cite{onyisi2023comparing}, and PC-based characterization of binding sites in viruses~\cite{springernatureInnophoresPoint} highlight the versatility of PCs. The illustrative visualization paradigm~\cite{viola2005tutorial} provides various techniques to enhance the interpretability of the various scientific visualizations~\cite{brambilla2012illustrative} using visual constructs such as textures, transparency, glyphs, etc. In practice, many visualization tools~\cite{li2019vapor,blender,QGIS_software} provide such functionalities to render and process PCs. However, many tools~\cite{li2019vapor,blender} fail to handle large amounts of high-resolution PC data. In this work, we choose the PC as a suitable representation for efficient scientific visualization at scale.

\subsection{Interactive High-Quality Visualization at Scale}
Owing to their reduced vertex buffer usage and low bandwidth requirements, point primitives have been a primary choice in real-time visualization of large scenes both in research~\cite{SCHUETZ-2021-PCC, kerbl3Dgaussians, nysjo2020raycaching} and practice (Potree~\cite{SCHUETZ-2020-MPC}, CloudCompare~\cite{CCsoftware}, QGIS~\cite{QGIS_software}, MegaMol~\cite{gralka2019megamol}). RayCaching~\cite{nysjo2020raycaching} is a point-based isosurface rendering method that caches points computed by ray intersections with scene elements and then renders through splatting. Compute Rasterizer (CR)~\cite{SCHUETZ-2021-PCC} is another high-performance PC renderer for massive datasets that uses compute shaders to avoid the use of conventional point primitive APIs (such as $\texttt{GL\_POINTS}$ in OpenGL and WebGL) and the accompanying inefficiencies of the standard graphics pipeline. This allows CR to improve the color quality, by reducing aliasing artifacts, and rendering speed, by proposing various optimizations on the buffers and vertex orders. Although these renderers are performant, they require explicit post-processing to render visual effects, which our method achieves via learning.




\subsection{Neural Point Cloud Rendering}
\label{sec:npbglit}
Point-based graphics has been a popular paradigm for visualization~\cite{kobbelt2004survey} using points~\cite{levoy1985use} and surfels/splats~\cite{zwicker2001surface, pfister2000surfels} as modeling primitives. More recently, the neural point-based graphics (NPBG)~\cite{aliev2020neural} approach, for rendering visual scenes reconstructed with images, extends the idea of neural deferred rendering~\cite{thies2019deferred} to PCs. Since then, many works~\cite{rakhimov2022npbg++, ruckert2022adop, li2023read} have expanded this idea for efficient and high-quality visual scene rendering. NPBG has been extended to eliminate per-point neural descriptor optimization via view-dependent prediction \cite{rakhimov2022npbg++}, and to incorporate differentiable camera parameters, such as exposure, white balance, and camera response function, for improved physical fidelity \cite{ruckert2022adop}. NPBG's capability has been extended to generate high quality renderings of larger visual scenes using an efficient sampling strategy of sparse PCs~\cite{li2023read}. Some high-quality neural PC rendering methods~\cite{dai2020neural, hu2023point2pix} also employ expensive neural frameworks, such as 3D CNNs and multiscale radiance fields, which are unsuitable for real-time rendering and post-processing of large PCs. All of these methods compute the PCs or equivalent neural representations from image-based inputs, typically using structure-from-motion or learning-based methods. Our work uses actual PCs, typically from scientific simulations or observations, as input, avoiding numerical inaccuracies in the PC data. Unlike natural scenes that rely on textural priors, scientific datasets consist of high-dimensional, multi-scalar attributes (e.g., pressure, velocity) that lack traditional image textures. These datasets often require semantic transformations, such as mapping to phase space, which creates a more complex optimization landscape for neural networks compared to natural image synthesis. NARVis addresses these challenges by enabling real-time, attribute-aware post-processing for the visualization of massive scientific point clouds.

%% file: sections/method.tex
\section{Neural Accelerated Renderer}


\begin{figure}[!ht]
    \centering
    \subfigure[Training Phase]{\includegraphics[width=0.95\linewidth]{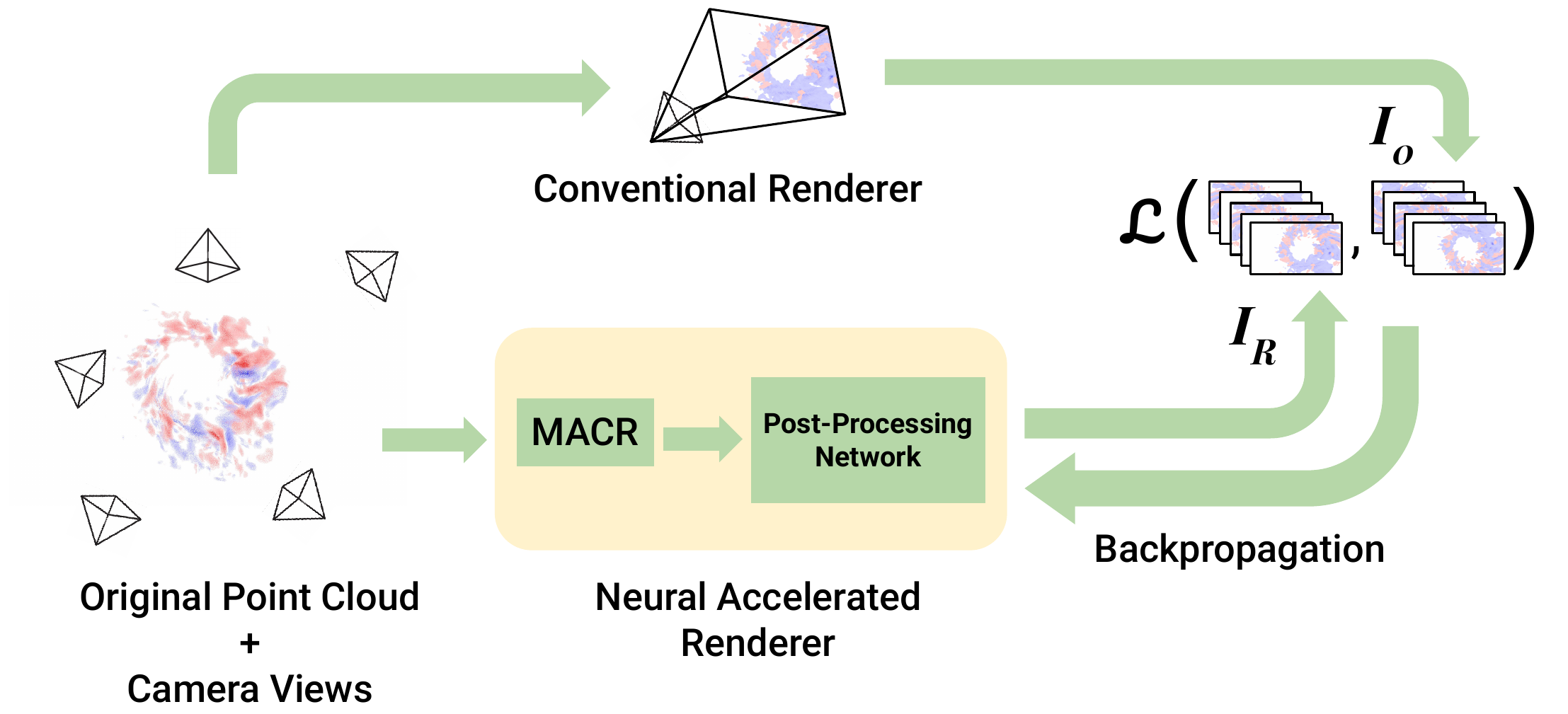}} 
    \subfigure[Inference Phase]{\includegraphics[width=0.95\linewidth]{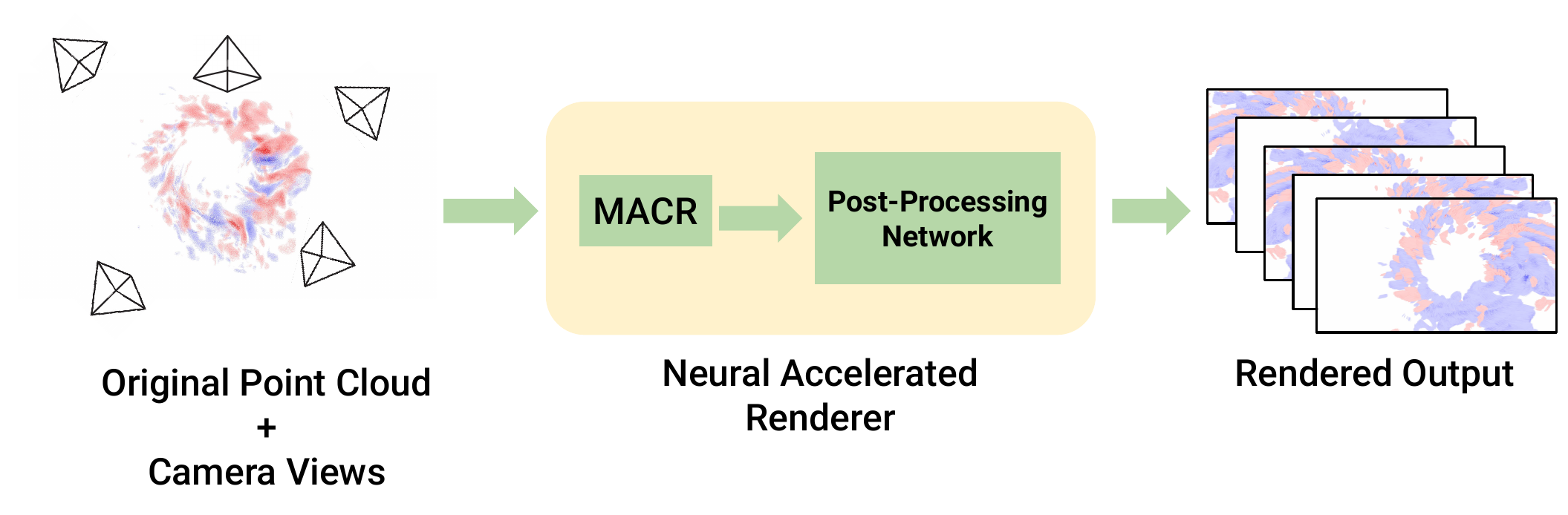}} 
    \caption{NARVis Overview. (a) We train NARVis with a large scientific PC data and a preferred renderer for learning geometry (PC configuration) and appearance (textures, color maps, etc. from the target visualizations) respectively. $I_O$ and $I_R$ represent the original rendering from the conventional renderer and NARVis respectively. (b) We only use the original (or modified) PC to render real-time high-quality renderings based on the user's viewing directions for inference.}
    \label{fig:overview}
\end{figure}

\subsection{Framework Overview}


In this work, we accelerate a given conventional renderer to visualize large-scale PCs in real-time. Specifically, we achieve this by learning how the view-space spatial distribution of points looks like given the color, depth, and additional point attributes that are rendered using a reference implementation of a high-quality glyph-based visualization technique. Figure \ref{fig:overview} shows an overview of the training and the inference phases of NARVis. NARVis has two major components - a high-performance multi-attribute compute rasterizer (MACR) and a post-processing network (see figure \ref{fig:nerual_rend} for more details). In the training phase, we provide flexibility to choose the high-quality renderer, as per the user's requirements, to create training datasets that NARVis learns to emulate. In the inference phase, MACR only needs to render the attributed points with depth testing from any position because the glyphs (geometry, orientation, texture, etc.) can be reconstructed for that view-space point/data distribution by the post-processing neural network in real-time. This also means that the post-processing network needs to be trained for each visualization style/glyph separately. However, if the learned view-space distributions are representative enough to cover the different possible visualization configurations, each trained network can render arbitrary datasets with the same emulated glyph or visualization technique. Thus, after the offline training phase, we can deploy NARVis to render high-quality visualizations of the original or even a modified PC (such as a subsampled one) in real-time. In this section, we present the details of each of the components of NARVis. We also provide additional details on the training data generation in section \ref{sec:train_data}.

\begin{figure}
    \centering
    \includegraphics[width=\linewidth]{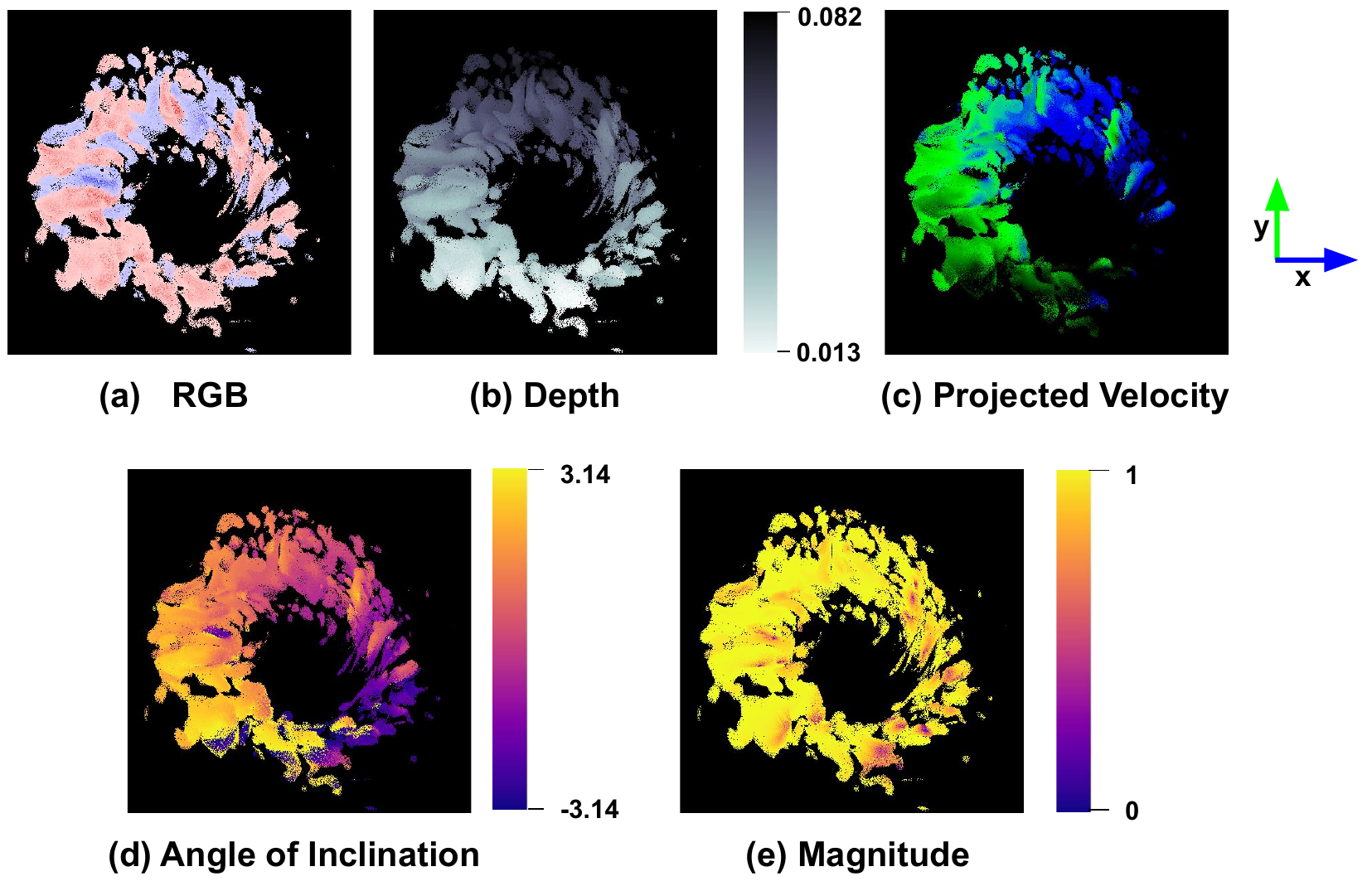}
    \caption{An example of multiple G-buffers rendered with MACR used in Hurricane visualization (RGB+D+Vel2D case from section \ref{sec:data_streams}). (a) RGB image of the point cloud, (b) the depth map, (c) projected 2D velocity (blue and green represent the magnitude of x and y components of the projected velocity vectors respectively), (d) angle of inclination of the projected velocity vector, and (e) the magnitude of the projected velocity vector. The choice of G-buffers to be rendered vary with the different visualization needs.}
    \label{fig:stream_viz}
\end{figure}

\subsection{Multi-Attribute Compute Rasterizer}

Unlike conventional rasterizers that render color and depth per pixel, we consider the various simulated or measured physical parameters to visualize the scientific PCs. In scientific visualization, such as flow field visualization, each field sample is a multi-attribute point. For example, the Lagrangian representation of hurricane and storm simulation data (refer to Figure \ref{fig:teaser}(a),(b)) contains field attributes such as atmospheric pressure, temperature, velocities, and pressure-altitude. We map some of these simulation parameters to the color channels and the 3D position coordinates using a user-defined transfer function. To accommodate more point attributes in the visualization, illustrative visualization methods~\cite{interrante2000harnessing} use glyphs and textures. To facilitate rendering such visualizations, we propose a multi-attribute compute rasterizer (MACR). MACR is a point-based rasterizer that renders PC points as pixels and can ingest other point field attributes, apart from the color and 3D position of the points, as additional information streams. The choice of the parameters to be visualized and the transfer function to be applied on the selected parameters rests with the users. We then use MACR to render multiple G-buffers for the scientific PC, each representing different simulated or measured physical parameters. These G-buffers serve as feature-rich inputs for the post-processing network in the next stage of NARVis, to learn the colors, positions, and glyphs and textures to be rendered around the point in the visualization.
\\

\begin{figure*}
    \centering
    \includegraphics[width=0.85\linewidth]{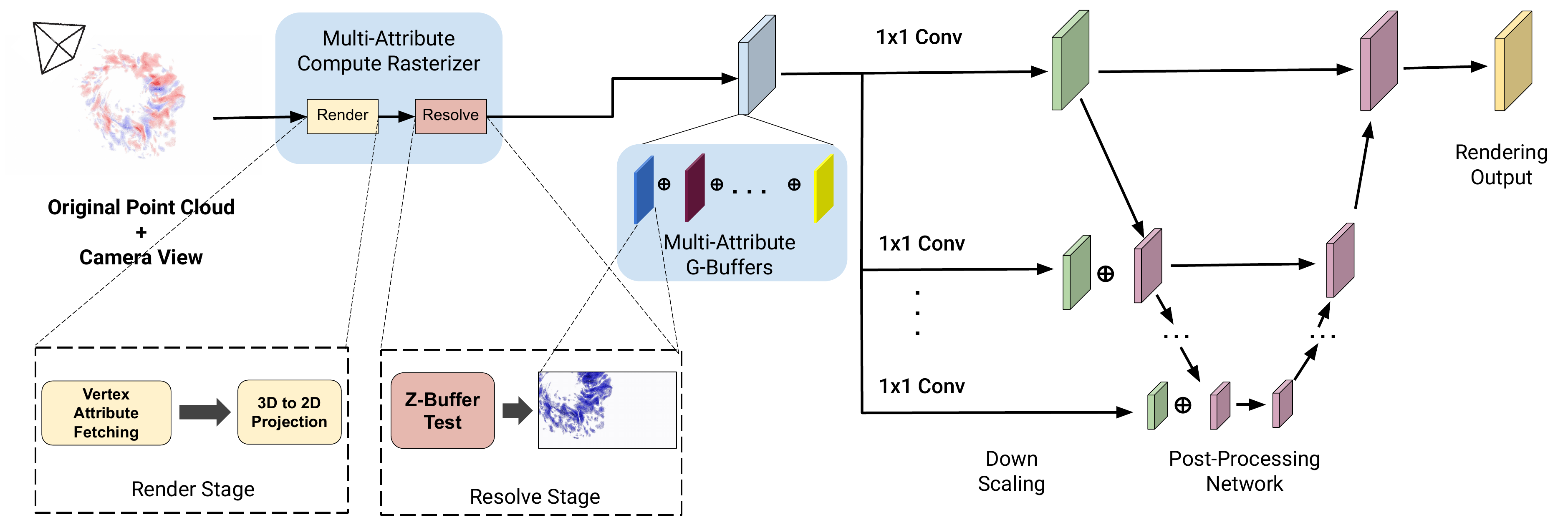}
    \caption{Details of the NARVis. We take the input view and the PC to output G-buffers for different point attributes of PC with an MACR, which processes each PC attribute in render and resolve stages (shown in the expanded view), and pass them as view-dependent multi-channeled features to our post-processing network, a U-Net~\cite{ronneberger2015u}, to generate final renderings. Before passing to the U-Net, we down scale and apply a \texttt{$1\times1$-Convolution} (retaining the number of channels) to the rendered G-buffers and concatenate ($\bigoplus$) them to the first U-Net blocks at the respective resolutions. (Green blocks: down scaled and \texttt{$1\times1$} convolved multiple G-buffers, Pink blocks: U-Net block outputs.)}
    \label{fig:nerual_rend}
\end{figure*}

\noindent We extend the capabilities of high-performance CR compute shaders~\cite{SCHUETZ-2021-PCC} with multi-attribute rasterization to process the multi-variate scientific data. We perform rendering in two stages - render and resolve (see figure \ref{fig:nerual_rend}). The render stage transforms the points from world to screen space and renders point depth to a custom framebuffer. While projecting $\sim$100M visible points onto a few million pixels, some of the pixels will have multiple points rendered to them due to the pigeonhole principle. Therefore, in the render stage, we also render the closest visible points with the lowest depth via early z-buffer testing to reduce rendering costs. The resolve stage finally transfers other corresponding point attributes to the final G-buffers. Unlike CR, we save on the computation cost by avoiding color averaging of the overlapping points. We can also process different point attributes based on the visualization needs with MACR. In this work, we use the velocity vector properties as additional point attributes for visualizing static vector fields and particle trajectory (see figure \ref{fig:stream_viz} and section \ref{sec:data_streams}). For rendering terrain data, we compute the projected point (splat) radius with MACR (see section \ref{sec:train_data}) and use it as an additional point attribute. Furthermore, we support different quantities of attributes (currently tested up to 8) and formats (\texttt{uint8}and \texttt{fp32})  for multi-variate data as per the visualization needs. 

\subsection{Post-Processing Network and Training Details}

We use a U-Net~\cite{ronneberger2015u} with gated convolution~\cite{yu2019free} as the post-processing network. The network takes the MACR rendered G-buffers, which render the point attributes of the visible points from a given input view, and concatenates them to create a view-dependent multi-channeled feature. We post-process the feature to generate multi-resolution feature (see Figure \ref{fig:nerual_rend}). Apart from rendering various visual effects, the benefit of a separate neural renderer (or a post-processing operator) is its invariance to the coordinate space conventions and transformations on the input 3D PC. Methods such as NPBG~\cite{aliev2020neural} effectively learn the local scene geometry and appearance with point-wise neural descriptors. However, this severely limits scalability to a large number of points. Instead, we use a 1x1 convolution layer to efficiently model the view-dependent local appearance around a point in the PC after projection to the image plane, removing the dependency on PC size. Also, we end up learning the view-dependent point descriptors of only the visible points due to MACR's z-buffer testing performed to only render the visible points from a given viewpoint. We empirically show that using a 1x1 convolution layer results in rendering quality comparable to the per-point neural descriptors while improving rendering speed (see \ref{sec:abl_comp}).
\\


\noindent We downsample the output of 1x1 convolution to create a set of multiscale images at different resolutions (specifically at 5 resolutions each downscaled by a factor of 0.5). This serves as a proxy to learn the implicit level of details (LoD) from the input PC. We then concatenate these multiscale images to the corresponding resolutions of interleaved first blocks of the U-Net encoder. Finally, we train the post-processing network with the following objective function:

\begin{multline}
\label{eq1}
    \mathcal{L}(I_R, I_O) = \mathcal{L}_{perc}(I_R, I_O) + \alpha\mathcal{L}_{reco}(I_R, I_O) + \beta \mathcal{L}_\nabla (I_R, I_O),
\end{multline}

\noindent where $I_R$ and $I_O$ are the NARVis output and the original images from the conventional renderer, respectively. We use perception loss~\cite{johnson2016perceptual}, $\mathcal{L}_{perc}$, for visual structure learning and the reconstruction loss (pixel-wise $L_2$ norm), $\mathcal{L}_{reco}$, for color consistency. For generating smooth rendering, we use the 2D total variation loss, $\mathcal{L}_\nabla$. $\alpha$ and $\beta$ are the optimization constants. 
\\


\noindent The encoder of post-processing U-Net employs 4 stages of downsampling, using gated convolutional blocks to handle sparse point data, while the decoder uses skip connections to fuse high-resolution features with feature maps upsampled with transpose convolution operations. The output of each gated convolution is passed through ELU activation function and $2\times2$ average pooling. Feature channels typically follow a $[64, 128, 256, 512, 1024]$ progression. In the objective function, we set the coefficients $\alpha$ and $\beta$ to $10^{-3}$ and $10^4$, respectively. We employ the Adam~\cite{kingma2014adam} optimizer with an initial learning rate of $10^{-3}$. We split the entire dataset into train and validation split of $9:1$. We pass the input training images at a resolution of $512 \times 512$. The training data is rendered with a field of view of $60^{\circ}$ and the near and far planes are placed at 0.1 and $2\times 10^5$ units. We train the network with a batch size of 16 for 100, 100, and 200 epochs for Hurricane, Storms, and Morro Bay, respectively (with $\sim2000$, $\sim4500$, and $\sim4500$ training samples respectively). It takes $\sim9$, $\sim20$, and $\sim38$ hours respectively to train on these datasets on an NVidia RTX A6000 GPU.


%% file: sections/experiments.tex
\section{Experiments and Results}

\subsection{Datasets and Evaluation Metrics}
We use two PCs - \textit{Hurricane} (408K points), based on the large eddy simulation of hurricanes~\cite{guimond2016impacts, guimond2018large}, and \textit{Storms} (45M points), produced from an atmospheric trajectory model using NASA GEOS winds~\cite{geos}. Datasets contain latitude and longitude coordinates of the points, their 3D velocity profiles, pressure, temperature and potential temperature, and pressure altitude fields from hurricane and storm simulations worldwide. We also test our framework on a significantly larger Morro Bay PC~\cite{schutz2022software} (350M points), a LIDAR scan of a large terrain area. These datasets represent different types of scientific 3D data with varied visualization requirements. Specifically, Hurricane represents a temporal snapshot of a vertical velocity scalar field, Storms is a simulation of particle trajectories under different atmospheric vector field influences, and Morro Bay is a photometric snapshot of a large 3D surface.
\\

\noindent We evaluate the rendering quality using peak signal-to-noise ratio (PSNR) and structural similarity index (SSIM) metrics and also report the per-frame latency and frame rates to measure the computational efficiency. We also report the memory footprint to measure the memory requirements of the methods. All the experiments use an NVidia GeForce RTX 2080 Ti GPU and RTX A6000 GPU (when the GPU and shared memory requirements exceed 13 GB).

\subsection{Training Data Generation}
\label{sec:train_data}

To generate the training dataset for NARVis, we utilized ParaView [5] to render 3D glyph-based visualizations, specifically employing elongated cuboids (or boxees) and cones to convey flow directionality (see GT of Figure \ref{fig:glyph}). To balance visual clarity with structural context, we rendered the underlying point cloud with high transparency while superimposing sparsified opaque glyphs sampled at every $50$th point (adding $\lfloor N/50\rfloor$ primitives). Critically, the standard rasterization of these geometric primitives in the ground-truth generator can introduce aliasing artifacts and high-frequency noise. Since NARVis is data-driven and designed to faithfully reconstruct the supervision signal, the network inherently learns to reproduce these rasterization artifacts, which manifest as the blur and spots observed in the final stylization results.
\\


\begin{figure}[!ht]
    \centering
    \begin{tabular}{cccc}
        \vspace{3mm}
        \rotatebox{90}{CR} & \includegraphics[width=.25\linewidth,valign=c]{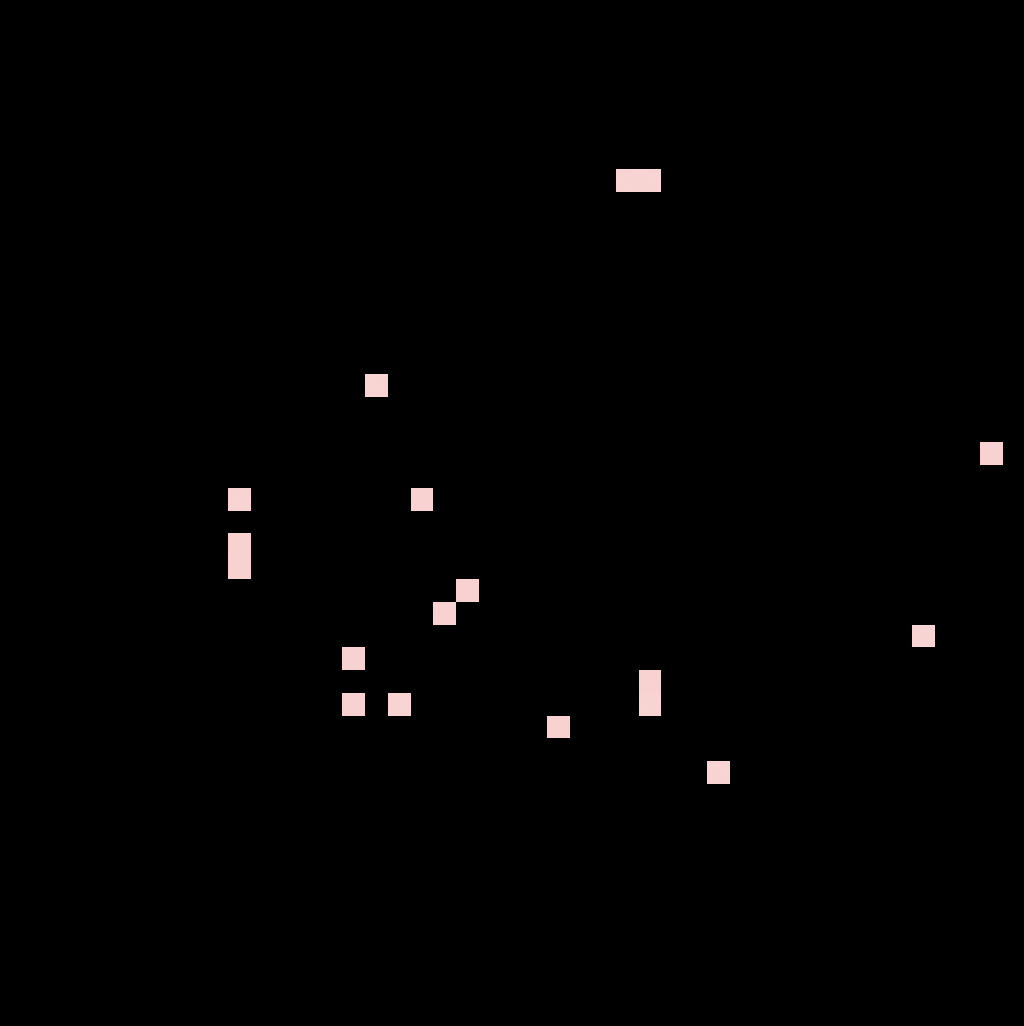} & \includegraphics[width=.25\linewidth,valign=c]{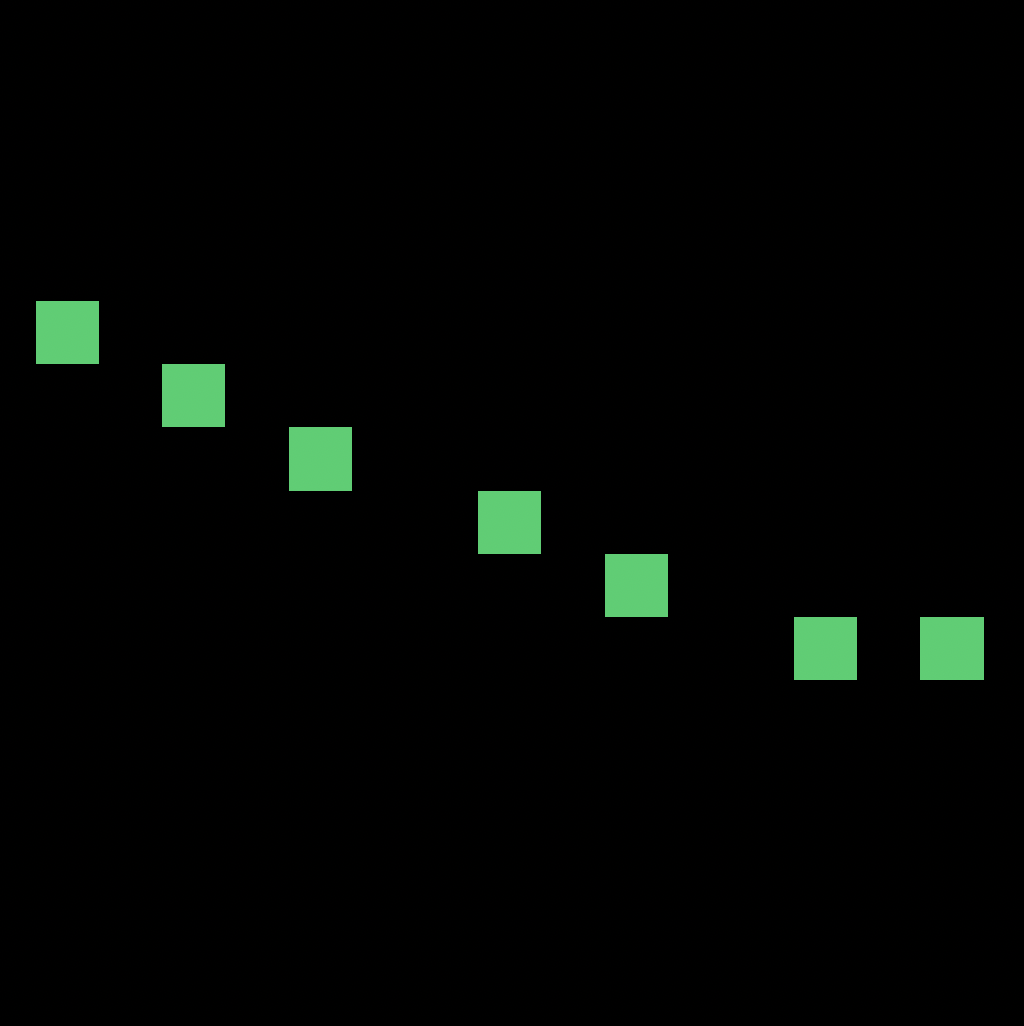} & \includegraphics[width=.25\linewidth,valign=c]{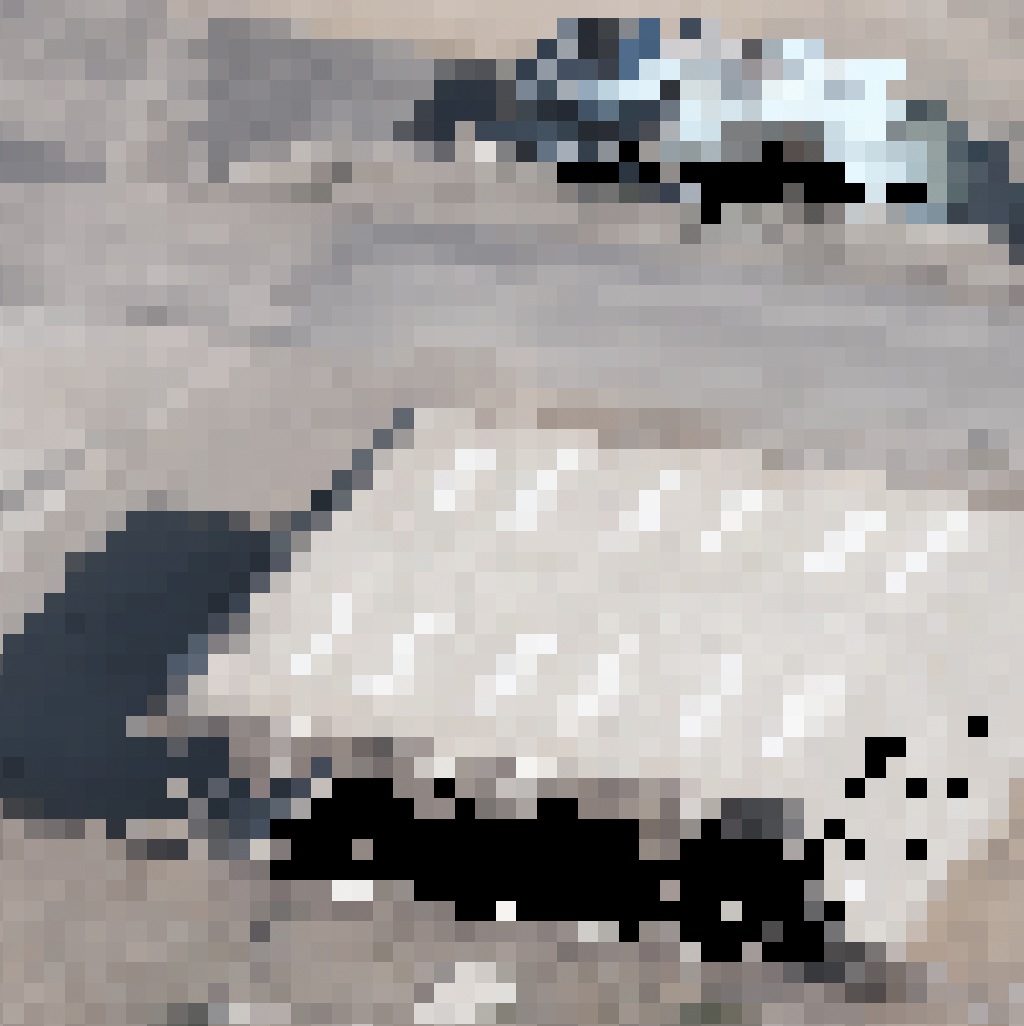}\\
        \vspace{3mm}
        \rotatebox{90}{GR} & \includegraphics[width=.25\linewidth,valign=c]{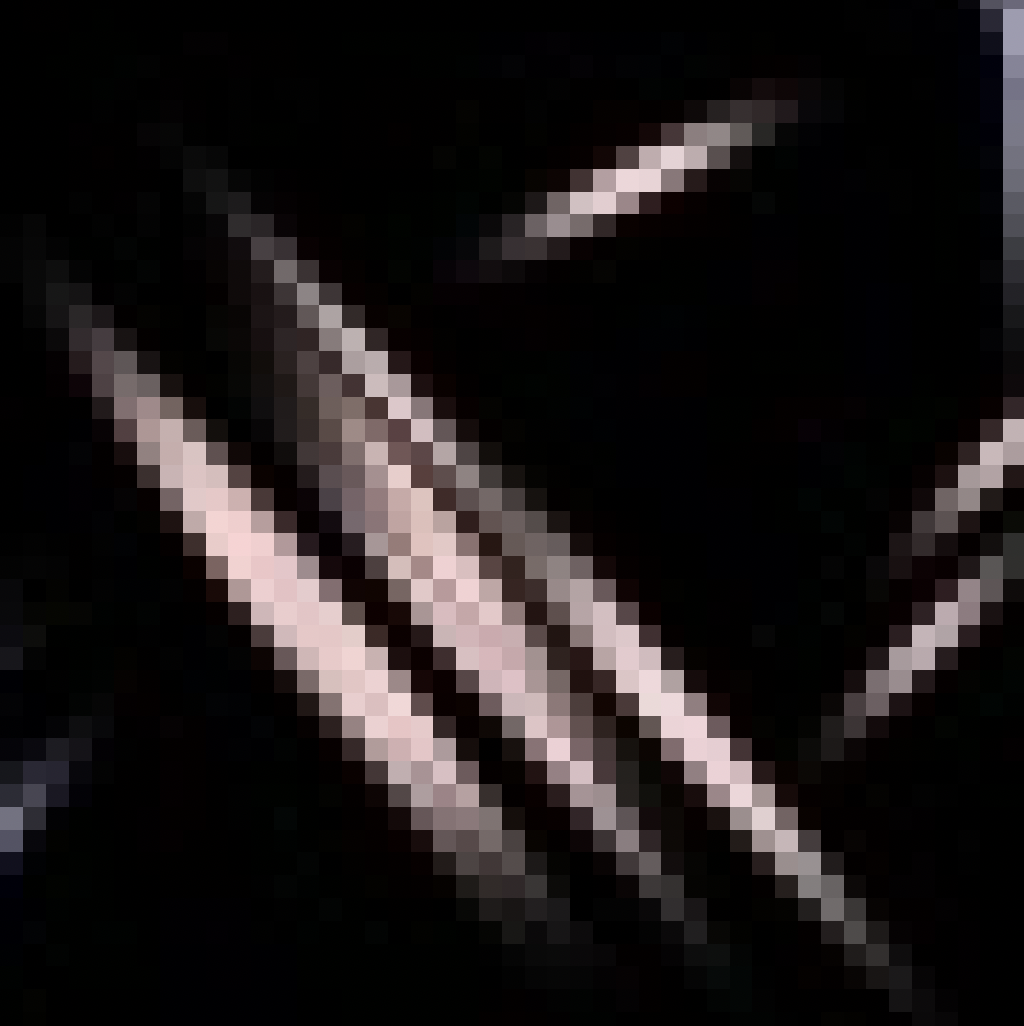} & \includegraphics[width=.25\linewidth,valign=c]{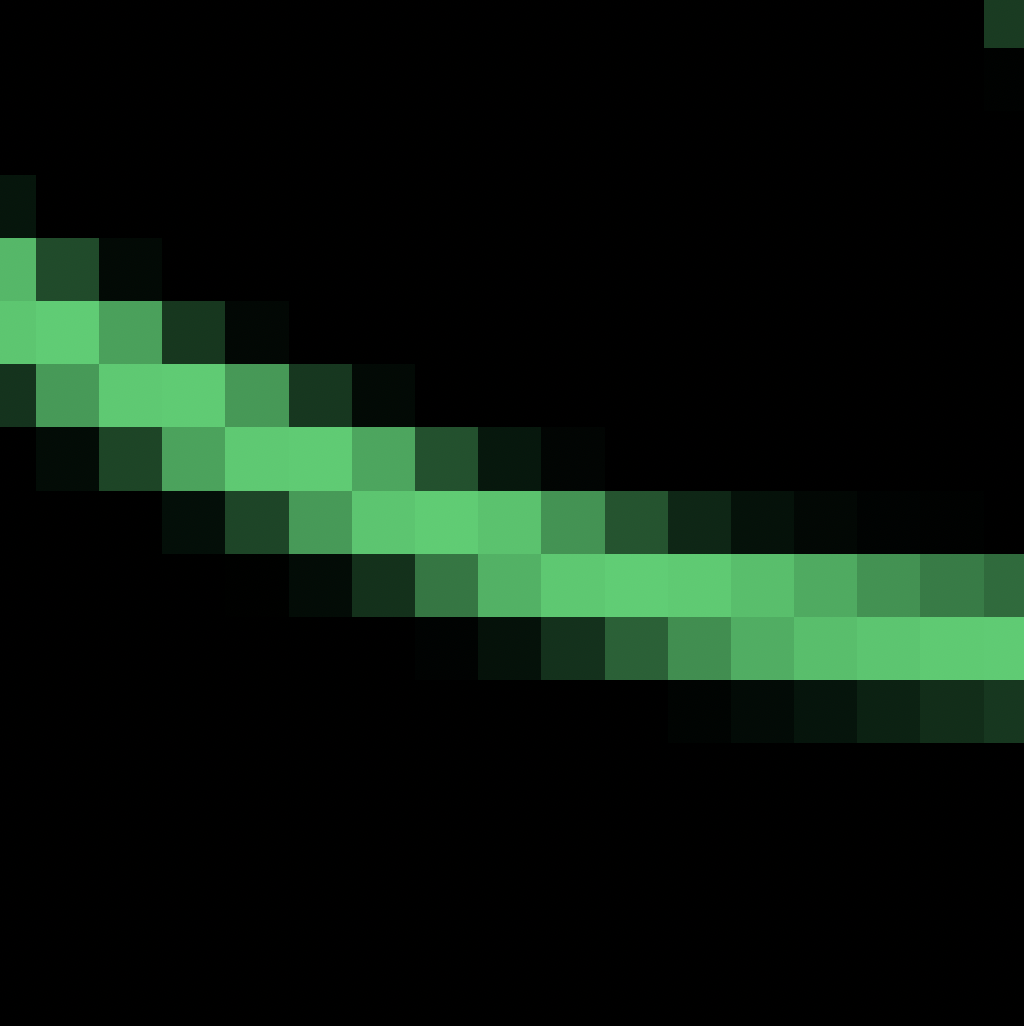} & \includegraphics[width=.25\linewidth,valign=c]{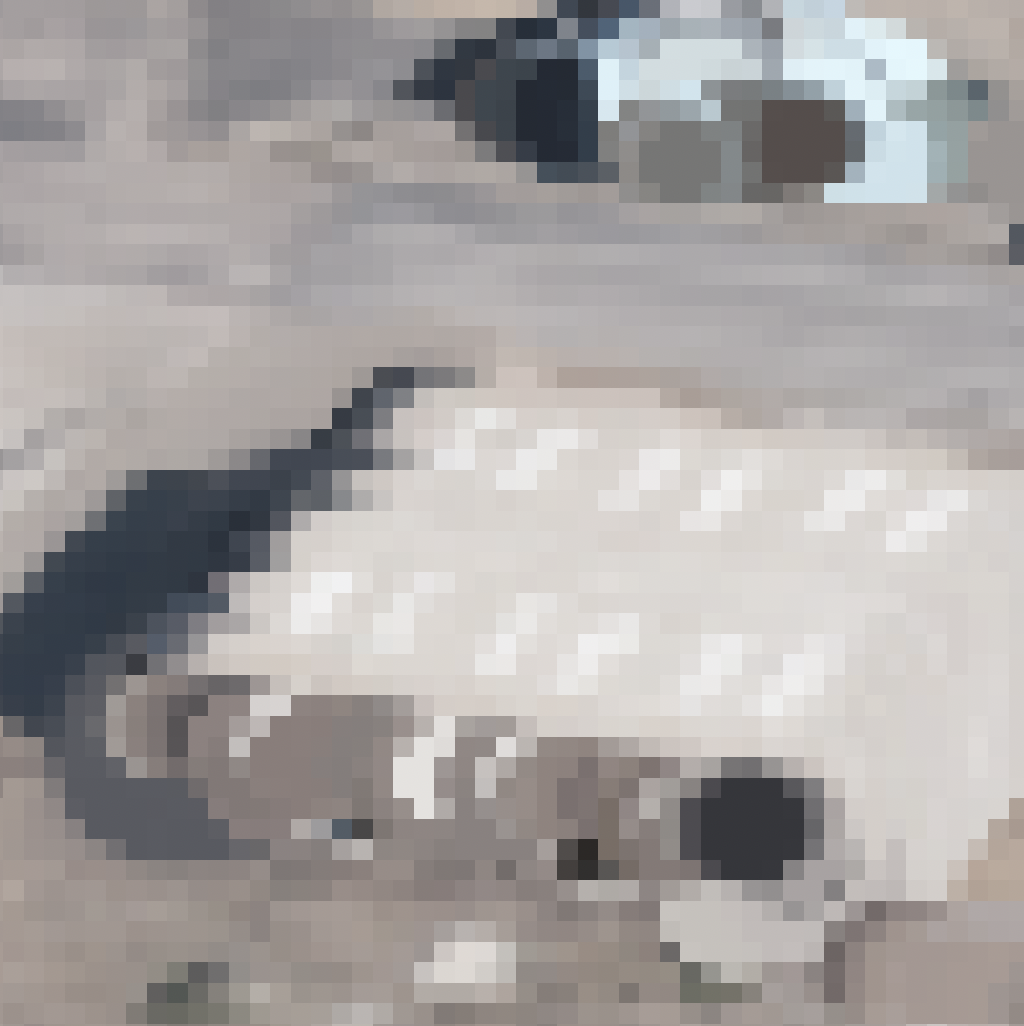}\\
        & (a) Hurricane & (b) Storms & (c) Morro Bay
    \end{tabular}
    \caption{Gaussian renderer's effects for PC visualization and training data creation.}
    \label{fig:gsfx}
\end{figure}

\noindent We also use a Gaussian Renderer (GR) for rendering various post-processing effects (see Figure \ref{fig:gsfx}) applicable for different tasks. GR only uses the forward renderer of the 3D Gaussian Splatting framework~\cite{kerbl3Dgaussians,zwicker2001surface} for post-processing point-based volumetric data. GR treats point samples as Gaussians which can capture the anisotropy in the data and also provide control over the geometry, color, and transparency of the point samples. For each point, $\mathbf{x}$, in the Hurricane and Storms, we include the point orientation to visualize the vector field, $\mathrm{F}: \mathbb{R}^3 \rightarrow \mathbb{R}^3$, by elongating Gaussians, $\mathcal{N}(\mathbf{x}, \Sigma)$, along the direction of the vector, $\mathbf{v} = \mathrm{F}(\mathbf{x})$. The covariance matrix, $\Sigma$, is computed as $\Sigma = RSR^T$, with $S = diag(s, 1, 1)$, (and generally the stretch factor $s >> 1$) as the scaling matrix and $R$ as the orthonormal rotation matrix whose dominant eigenvector is $\mathbf{v}$. We use a stretch factor, $s$, of 300 and 200 for Hurricane and Storms, respectively. For the Morro Bay dataset, we also control the splat sizes by scaling the Gaussian covariance matrices isotropically, which is useful for filling holes in the image. For hole filling, we set the splat radius as the average distance between a point and its four nearest neighbors. Furthermore, we carefully select viewpoints to cover all the scene elements (refer to the supplementary material).

\subsection{Comparative Evaluation}
\label{sec:compare}
\begin{figure*}[!ht]
    \centering
    \begin{tabular}{cccccc}
        \vspace{3mm}
        \rotatebox[origin=c]{90}{Hurricane} & \raisebox{-0.5\height}{\includegraphics[width=.15\linewidth]{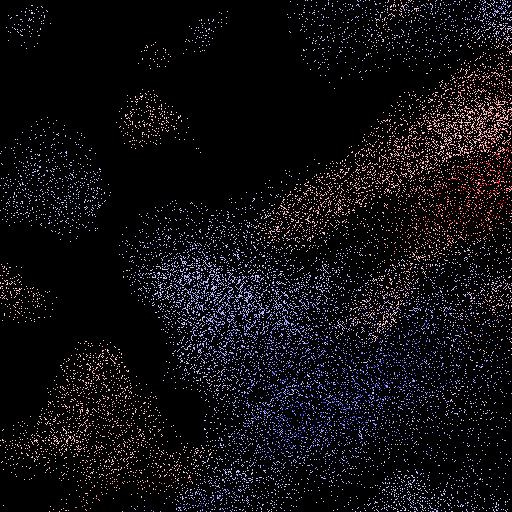}} & \raisebox{-0.5\height}{\includegraphics[width=.15\linewidth]{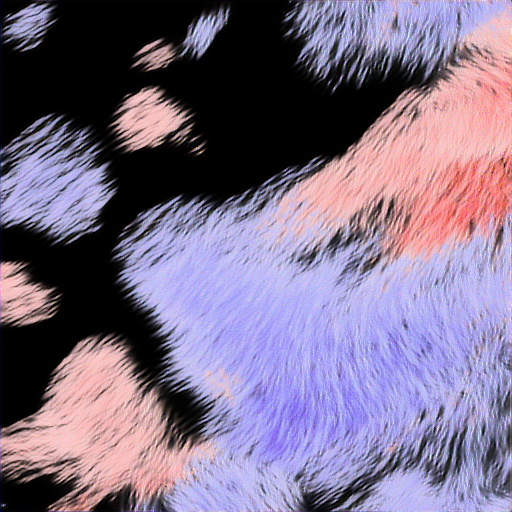}} & \raisebox{-0.5\height}{\includegraphics[width=.15\linewidth]{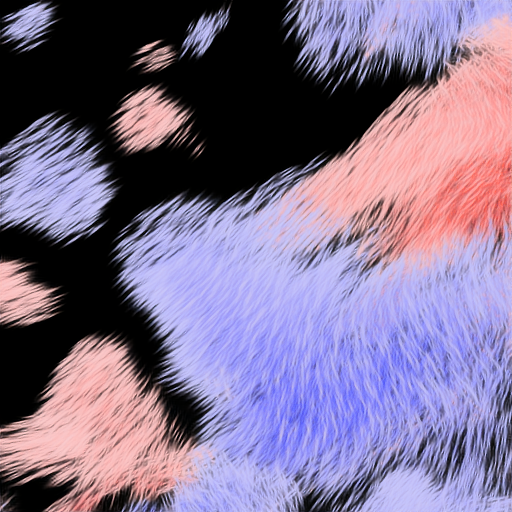}} & \raisebox{-0.5\height}{\includegraphics[width=.15\linewidth]{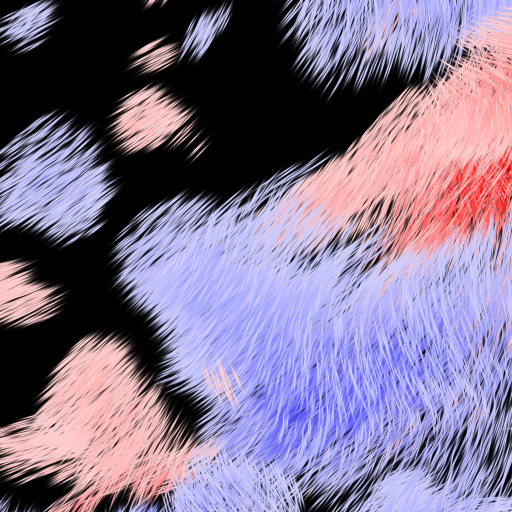}} & \raisebox{-0.5\height}{\includegraphics[width=.15\linewidth]{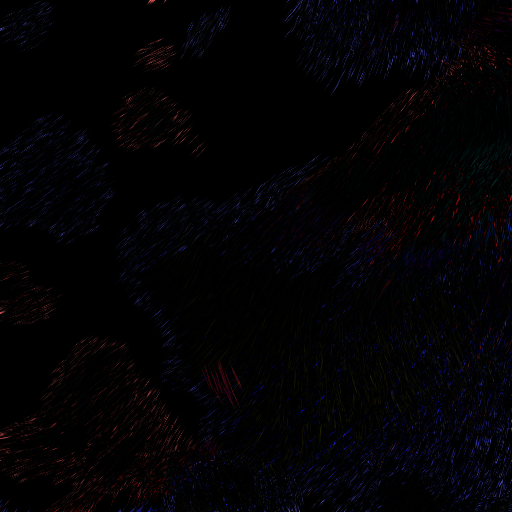}}\\
        \vspace{3mm}
        \rotatebox[origin=c]{90}{Storms} & \raisebox{-0.5\height}{\includegraphics[width=.15\linewidth]{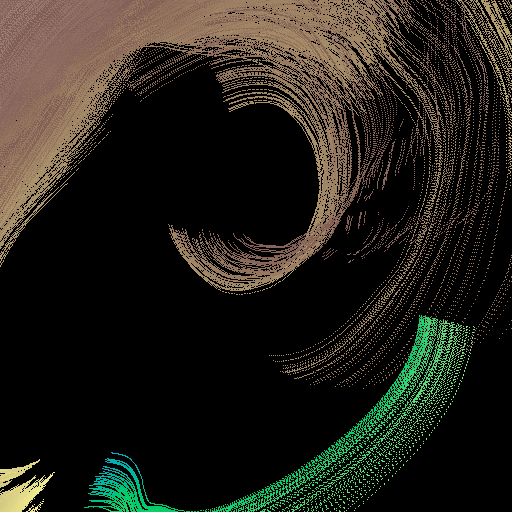}} & \raisebox{-0.5\height}{\includegraphics[width=.15\linewidth]{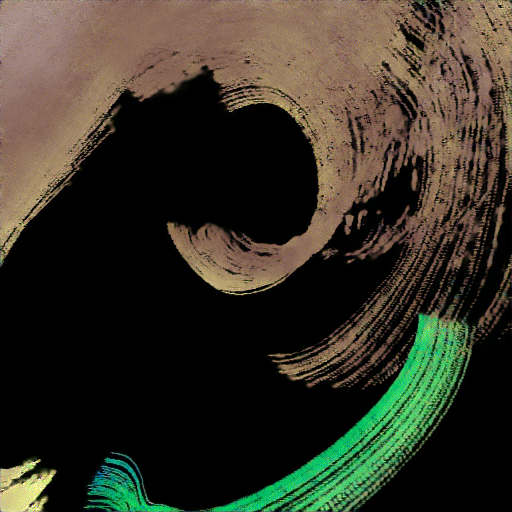}} & \raisebox{-0.5\height}{\includegraphics[width=.15\linewidth]{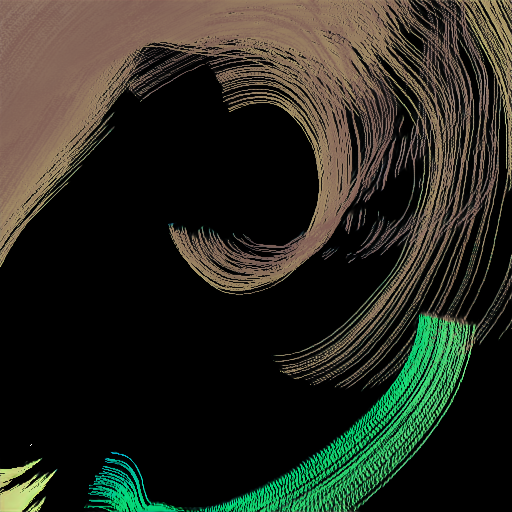}} & \raisebox{-0.5\height}{\includegraphics[width=.15\linewidth]{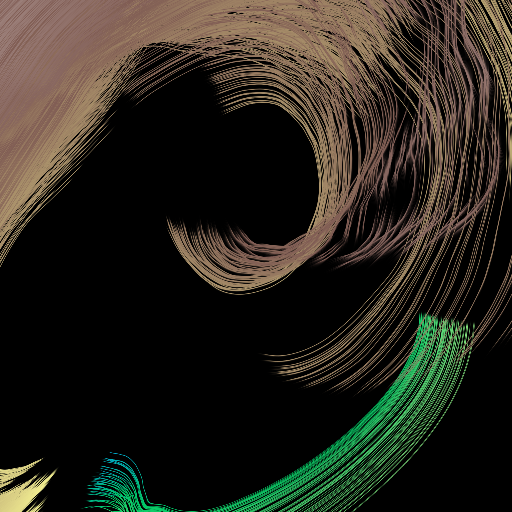}} & \raisebox{-0.5\height}{\includegraphics[width=.15\linewidth]{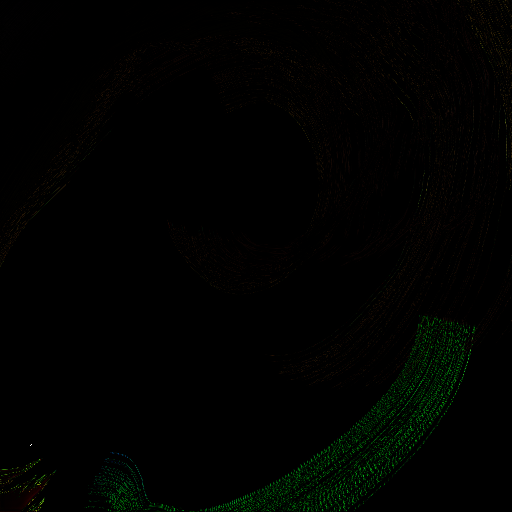}}\\
        \vspace{3mm}
        \rotatebox[origin=c]{90}{Morro Bay$^\dagger$} & \raisebox{-0.5\height}{\includegraphics[width=.15\linewidth]{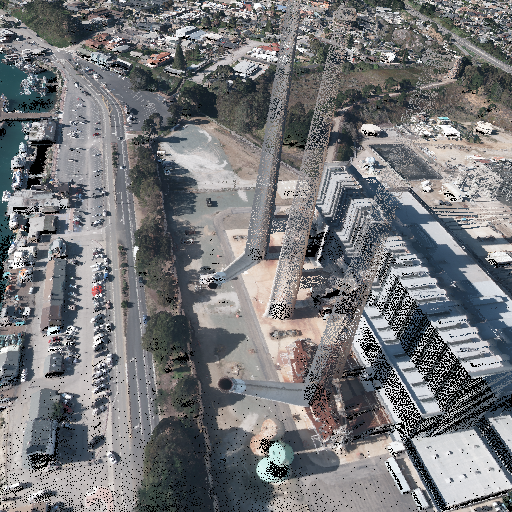}} & \raisebox{-0.5\height}{\includegraphics[width=.15\linewidth]{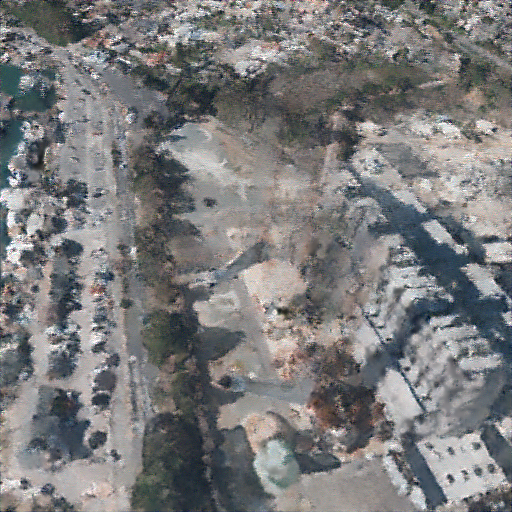}} & \raisebox{-0.5\height}{\includegraphics[width=.15\linewidth]{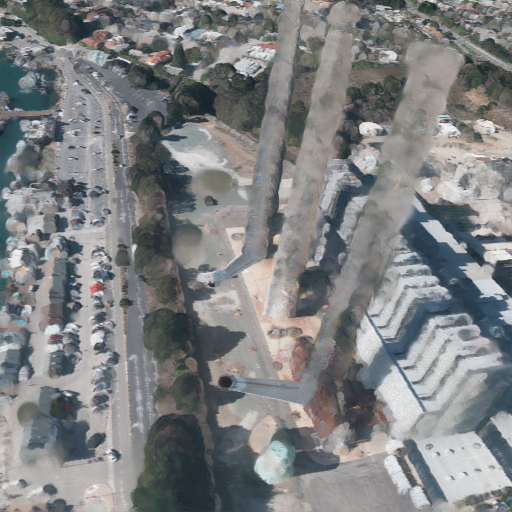}} & \raisebox{-0.5\height}{\includegraphics[width=.15\linewidth]{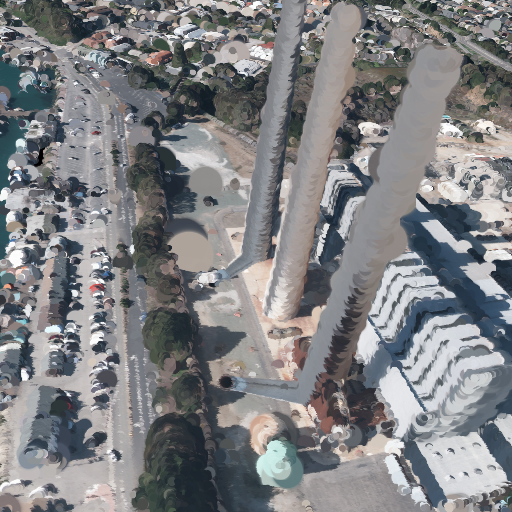}} & \raisebox{-0.5\height}{\includegraphics[width=.15\linewidth]{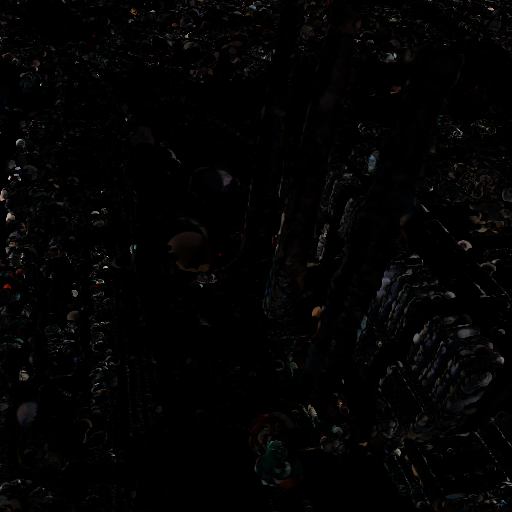}}\\
        & (a) CR & (b) NPBG & (c) NARVis & (d) GR & (e) Diff
    \end{tabular}
    \caption{Qualitative comparison of different renderers. NPBG and NARVis are trained on GR renderings (the ground truth). Diff column (e) shows the squared error between the NARVis and GR images. $^\dagger$GR and NPBG used 0.5× and 0.1× the original number of MorroBay points, respectively, due to memory constraints. More results in the supplementary materials.}
    \label{fig:comp}
\end{figure*}

\begin{figure}[!ht]
    \centering
    \begin{tabular}{cccc}
        \rotatebox[origin=c]{90}{Boxes} & \includegraphics[width=.25\linewidth,valign=c]{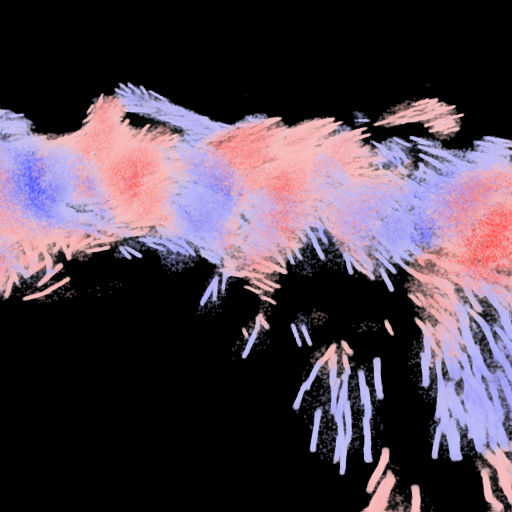} & \includegraphics[width=.25\linewidth,valign=c]{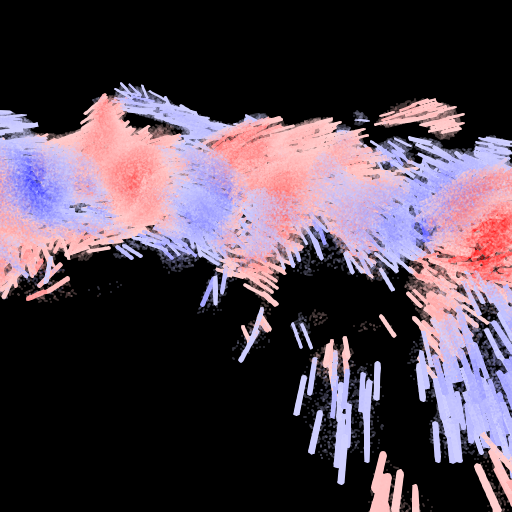} & \includegraphics[width=.25\linewidth,valign=c]{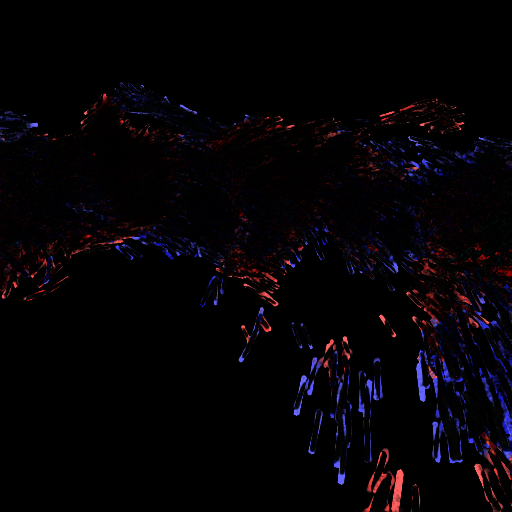}
        \vspace{3mm} \\
        \rotatebox[origin=c]{90}{Cones} & \includegraphics[width=.25\linewidth,valign=c]{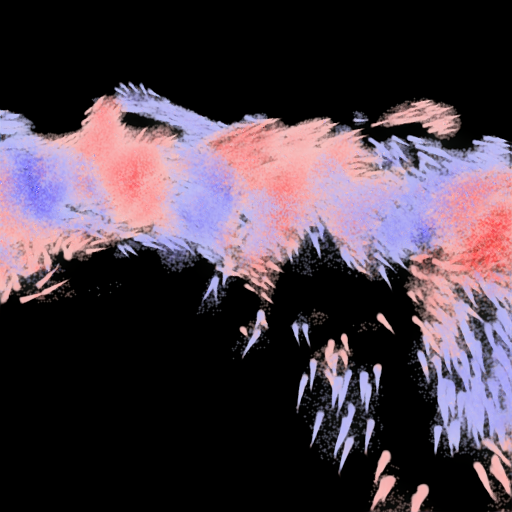} & \includegraphics[width=.25\linewidth,valign=c]{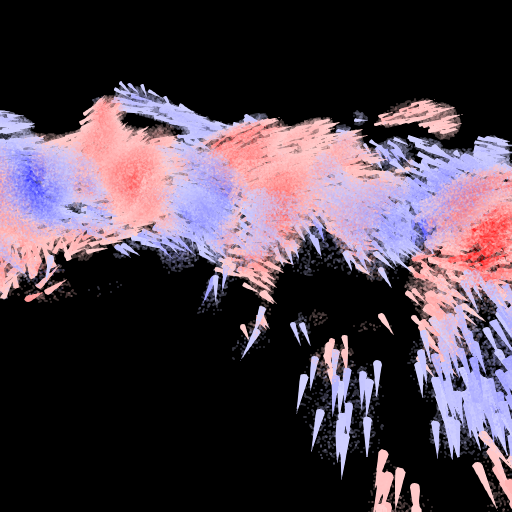} & \includegraphics[width=.25\linewidth,valign=c]{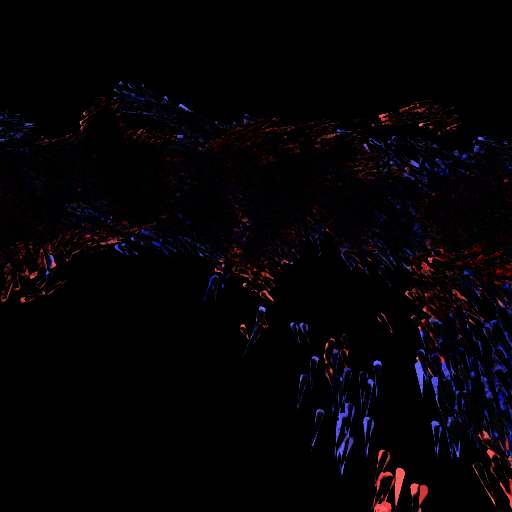}
        \vspace{3mm} \\
        \rotatebox[origin=c]{90}{Gaussians} & \includegraphics[width=.25\linewidth,valign=c]{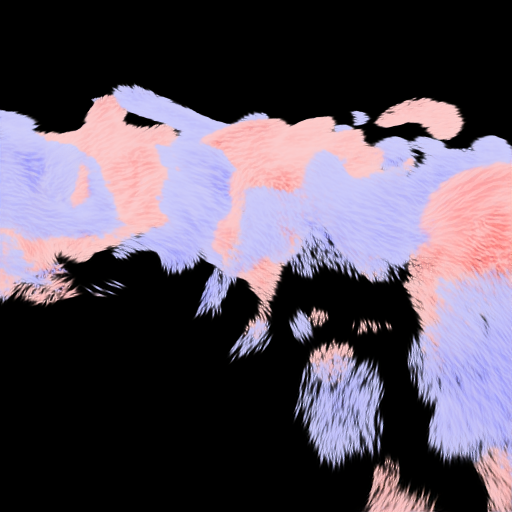} & \includegraphics[width=.25\linewidth,valign=c]{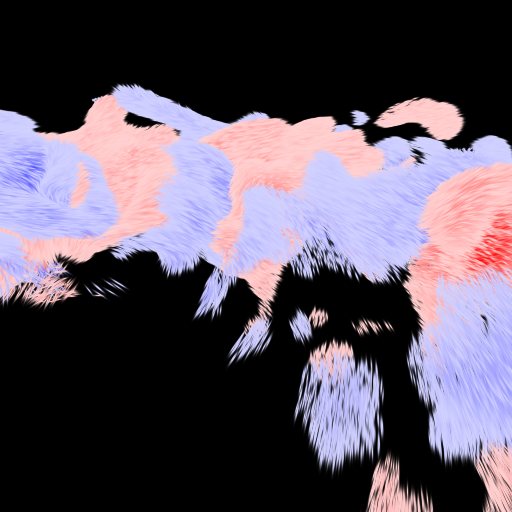} & \includegraphics[width=.25\linewidth,valign=c]{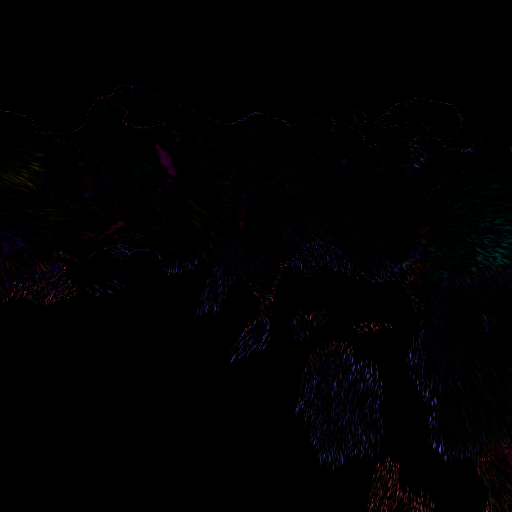}
        \vspace{3mm} \\
        & (a) NARVis & (b) GT & (c) Diff
    \end{tabular}
    \caption{NARVis Supports Glyph Stylization. NARVis outputs (a) are adaptive to different glyph style requirements, that are useful for illustrative visualization, such as elongated cuboids (boxes), cones, and Gaussians. Ground truth (GT) (b) are rendered with GR for Gaussians and with Paraview for other glyphs. We also visualize the squared error (c) Diff between (a) and (b). }
    \label{fig:glyph}
\end{figure}

\noindent \textbf{Rendering Performance:} We compare our method's performance, specifically latency, memory footprint, and rendering quality, against the following state-of-the-art renderers, (in Tables \ref{tab:comparison}, \ref{tab:comp_glyph} and Figures \ref{fig:comp}, \ref{fig:glyph}):

\begin{itemize}
    \item \textbf{Gaussian Renderer (GR)}~\cite{zwicker2001surface,kerbl3Dgaussians}: This forward renderer rasterizes Gaussians for each point in the PC allowing easy manipulation of the point geometry with high-quality rendering. GR is a naive implementation of only the forward renderer of 3DGS~\cite{kerbl3Dgaussians} differentiable rendering pipeline.
    \item \textbf{Compute Rasterizer (CR)}~\cite{SCHUETZ-2021-PCC}: We compare the early-z buffer tests CR variant without vertex order modification for Hurricane and Storms data and use Morton-ordered Morro Bay data. CR is optimized for both low-latency and high-quality colors mitigating anti-aliasing artifacts.
    \item \textbf{Paraview}~\cite{ayachit2015paraview}: This is a multi-platform data analysis and visualization toolkit supporting scripting-based glyph visualization. Paraview also supports interactive visualization by using a decimated PC for interaction and renders full resolution PC for static visualization. We compare against Paraview for glyph-based visualizations in our work.
    \item \textbf{NPBG}~\cite{aliev2020neural}: This neural PC rendering method uses a simple OpenGL renderer with learnable point descriptors.
\end{itemize}

\noindent NARVis balances rendering latency and quality and outperforms the other renderers in rendering quality across all the datasets. It is only slower than high-performance CR which has lower visual fidelity than NARVis, especially in sparser regions, as it renders points as pixels. GR has the highest image quality as it is our reference renderer. However, GR also has a high rendering latency due to the increased $\alpha$-blending overheads (involving sorting) in large PCs. NPBG, an NN based method, has better rendering quality than CR but is memory intensive as it uses per-point descriptors. NPBG also has a higher latency as its OpenGL rasterizer uses $\texttt{GL\_POINTS}$ primitive to render frames at multiple image resolutions, making it unscalable for real-time rendering of large full-resolution PCs. For small PCs, G-buffer overhead exceeds the per-point storage costs of NPBG and GR and using sparse data structures could improve the memory footprints.\\

\noindent \textbf{Glyph Stylization Support:} NARVis adapts to different glyph-based visualization styles. We observe that NARVis renders the glyph visualizations $>2\times$ faster than Paraview (see Table \ref{tab:comp_glyph}). This visualization is particularly challenging as it involves a transparent PC component that can occlude some glyph points. As the occluded points are ignored in MACR from the Z-buffer test, some glyph points are not learned by the post-processing NN and, thus, are not rendered in the final visualization. To overcome this limitation, we render the point corresponding to the frontmost visible glyph by ignoring the non-glyph points in front of it. This helps NARVis to learn visually high-fidelity renderings (refer to Figure \ref{fig:glyph} and \textit{VisGlyph} entries in Table \ref{tab:comp_glyph}).

\begin{table}[!ht]
  \centering
  \begin{tabular}{%
  	  r%
  	  	*{5}{c}%
  	}
    \toprule
        Method & \rotatebox{0}{PSNR} &   \rotatebox{0}{SSIM} & \makecell{Latency \\(in ms) / fps} & \makecell{Memory \\ (in GB)} \\
        \midrule
        \multicolumn{5}{l}{\textit{Hurricane (408K points)}} \\
        \midrule
        CR &  9.57 & 0.623 & \textbf{1.01 / 989.84} &  0.80 \\
        GR &  - &  - &  28.99 / 34.49 &  \textbf{0.40} \\
        NPBG &  22.76 & 0.788 &  13.64 / 73.32 &  1.00 \\
        Ours & \textbf{24.65} & \textbf{0.813} & 6.97 / 143.44 & 1.70 \\
        \midrule
        \multicolumn{5}{l}{\textit{Storms (45M points)}} \\
        \midrule
        CR &  17.92 & 0.707 & \textbf{1.03 / 966.83} &  \textbf{1.13} \\
        GR &  - &  - &  3515 / 0.28 &  7.10 \\
        NPBG &  23.13 & 0.834 &  80.07 / 12.49 &  4.90 \\
        Ours & \textbf{28.02} & \textbf{0.915} &  6.58 / 152.01 & 3.10 \\
        \midrule
        \multicolumn{5}{l}{\textit{Morro Bay (350M points)}} \\
        \midrule
        CR & 17.28 & 0.312 & \textbf{1.05 / 951.09} & \textbf{5.79} \\
        GR$^{*}$ &  - &  - &  5536 / 0.18 &  23.93 \\
        NPBG$^{\dagger}$ & 23.76 & 0.683 &  64.82 / 15.428 & 5.98 \\
        Ours & \textbf{24.49} & \textbf{0.683} &  7.91 / 126.38 &  12.73 \\
        
  \bottomrule
  \end{tabular}
  \caption{%
  	Comparison of rendering quality (w.r.t GR rendering), latency, and memory footprints of different renderers. Note: we use the original PC resolution for rendering. (\textbf{$\mathbf{^{*}}$used $\mathbf{0.5\times}$ and $\mathbf{^{\dagger}}$ used $\mathbf{0.1\times}$ the number of points due to memory restrictions})%
  }
  \label{tab:comparison}
\end{table}

\begin{table}[!ht]
  \centering
  \begin{tabular}{%
  	  r%
  	  	*{5}{c}%
  	}
    \toprule
        Method & \rotatebox{0}{PSNR} &   \rotatebox{0}{SSIM} & \makecell{Latency \\(in ms) / fps} & \makecell{Memory \\ (in GB)} \\
        \midrule
        Paraview & - & - & 16.71 / 59.85 & \textbf{0.45}\\
        Ours (Box) & 18.06 & 0.778 & 7.26 / 137.63 & 1.70 \\
        \makecell{Ours (Box) +\\VisGlyph} & 20.89 & \textbf{0.779} & 7.27 / 137.55 & 1.70 \\
        Ours (Cone) & 18.73 & 0.745 & \textbf{7.25 / 137.84} & 1.70 \\
        \makecell{Ours (Cone) +\\VisGlyph} & \textbf{21.14} & 0.731 & 7.26 / 137.74 & 1.70 \\

  \bottomrule
  \end{tabular}
  \caption{%
  	Comparison of rendering quality (w.r.t Paraview rendering), latency, and memory footprints under different glyph stylizations for Hurricane Glyph dataset (with 417K points).%
  }
  \label{tab:comp_glyph}
\end{table}

\subsection{Generalizability of NARVis}
\label{sec:general}
\noindent We explore the generalizability aspects of NARVis to show its broader applicability. To this end, we train NARVis on a source PC to visualize a different target PC with similar visualization requirements.

\subsubsection{Training on Partial Dataset}
\begin{figure}[!ht]
  \centering
  \begin{tabular}{ccc}
  \vspace{3mm}
  \includegraphics[width=.29\linewidth,valign=c]{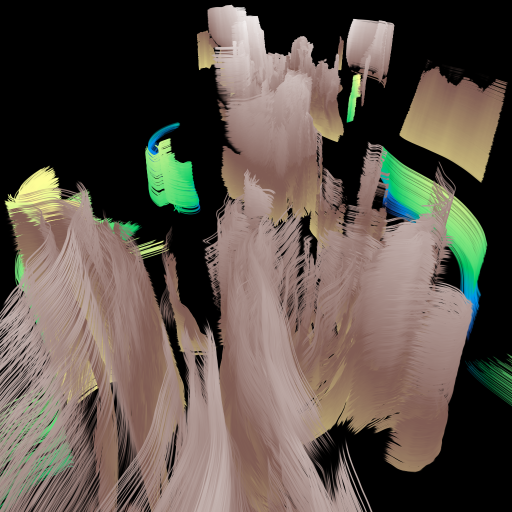} & \includegraphics[width=.29\linewidth,valign=c]{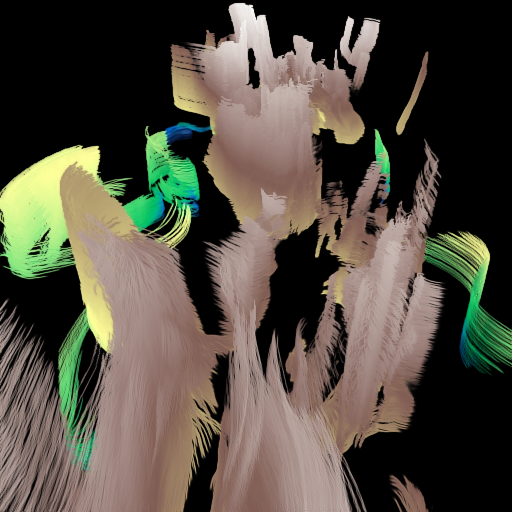} & \includegraphics[width=.29\linewidth,valign=c]{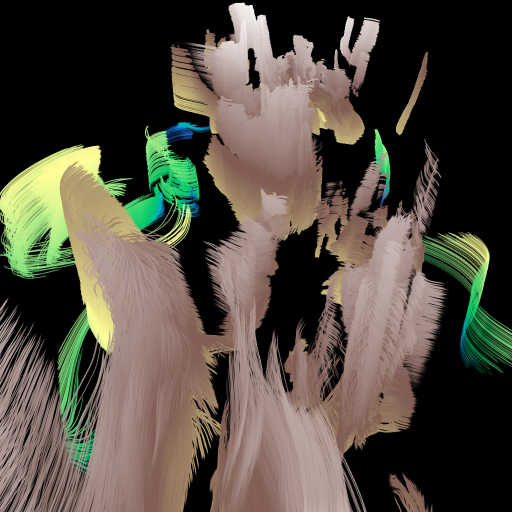}\\
  \vspace{3mm}
  \includegraphics[width=.29\linewidth,valign=c]{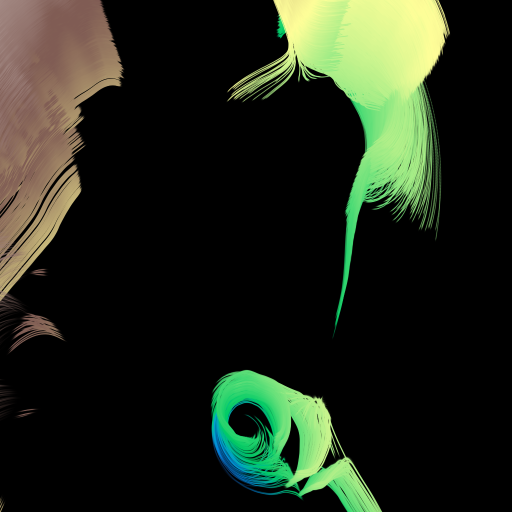} & \includegraphics[width=.29\linewidth,valign=c]{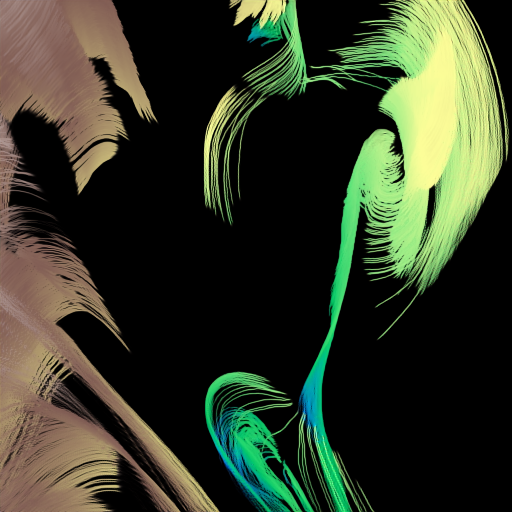} & \includegraphics[width=.29\linewidth,valign=c]{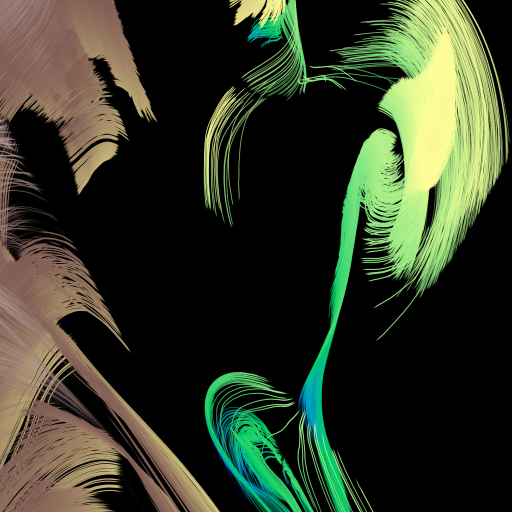}\\
  \vspace{3mm}
  \includegraphics[width=.29\linewidth,valign=c]{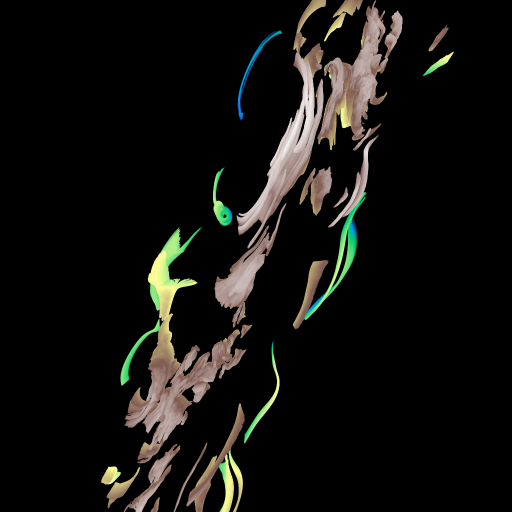} & \includegraphics[width=.29\linewidth,valign=c]{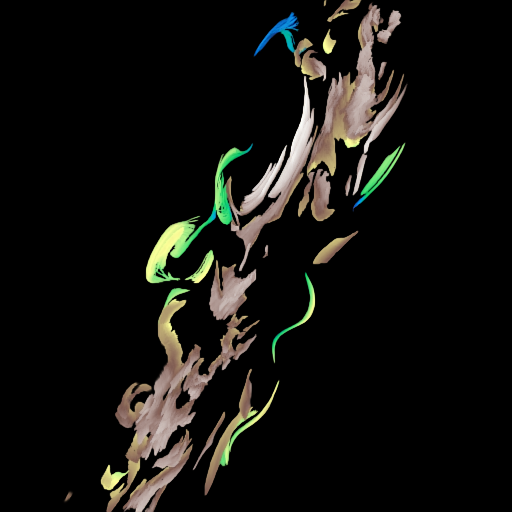} & \includegraphics[width=.29\linewidth,valign=c]{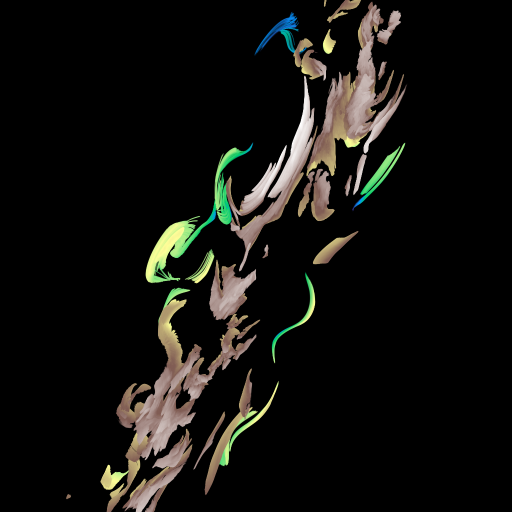}\\
   (a) Source PC & (b) Target PC & (c) GR
  \end{tabular}
  \caption{Generalizability of NARVis Trained Under Partial Storms Dataset. NARVis learns the post-processing effects from the source PC, (a), for visualizing the target PC, (b), irrespective of diverse PC geometry and properties. (c) GR renderings from the corresponding views for reference.}
  \label{fig:genexp}
\end{figure}

We demonstrate the generalizability of NARVis within a given dataset when trained on partial PC. For this experiment, we split the Storms dataset by time slices of the trajectories. Of the 301 time slices, we use the first 200 time slices of trajectory for the source PC and the rest 101 slices form the target PC. We also increase the viewpoint grid resolution (refer to supplementary materials) to $15 \times 15$, obtaining more training views. We achieve an average PSNR of 30.23 and an average SSIM of 0.903 (w.r.t. GR) for the target PC. Figure \ref{fig:genexp} shows the NARVis rendering of the source and target PCs from the same views. Unlike NPBG~\cite{aliev2020neural}, which cannot render a variable number of points during inference, NARVis effectively learns the required post-processing effects irrespective of the specific properties of the source/target PC.

\subsubsection{Generalizability Across Datasets}
\begin{figure}[!ht]
  \centering
  \begin{tabular}{cc}
  \multicolumn{2}{c}{\includegraphics[width=0.85\linewidth,valign=c]{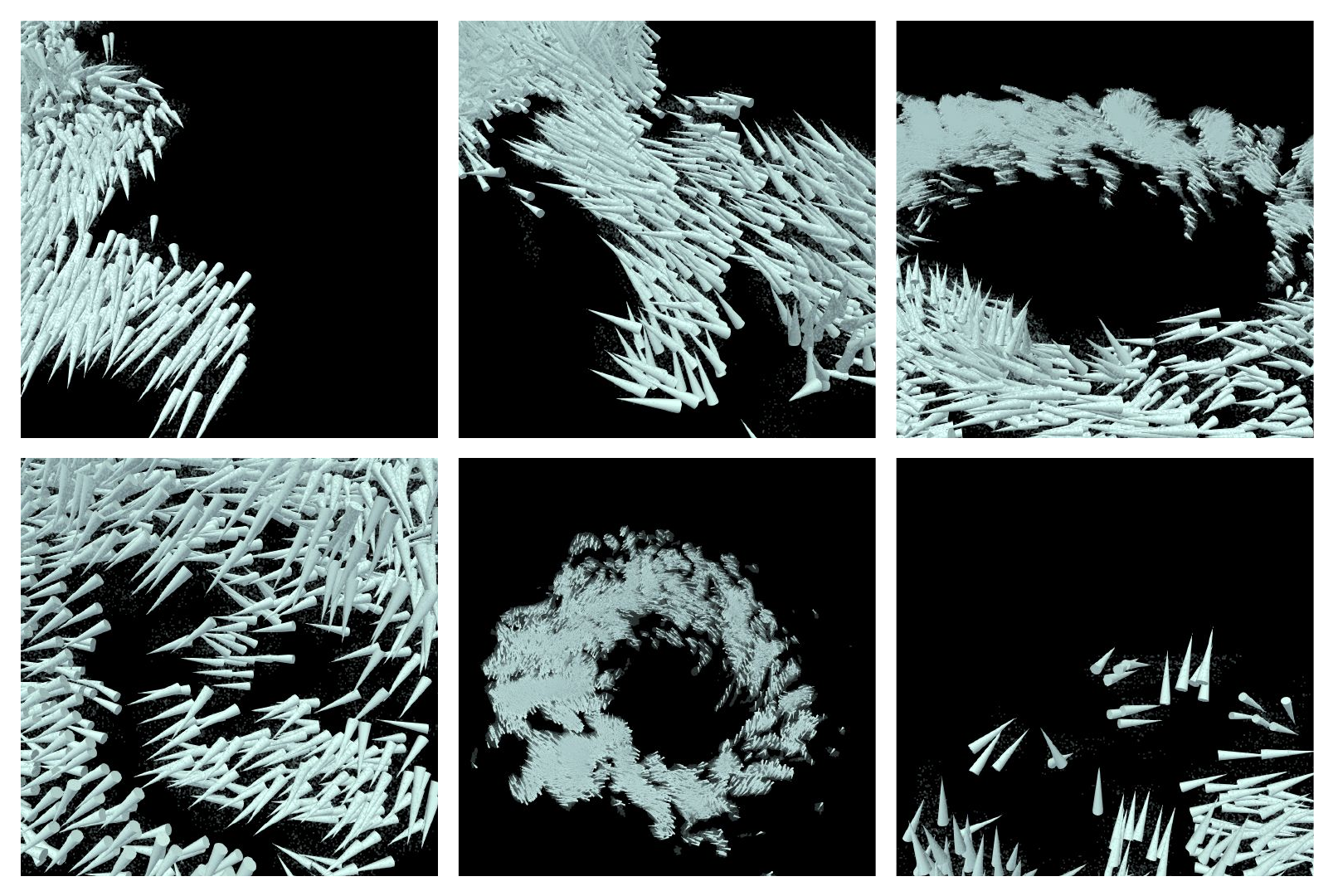}}
  \vspace{3mm}\\
  \multicolumn{2}{c}{(a) Training Samples from Source PC}
  \vspace{3mm}\\
  \includegraphics[width=.38\linewidth,valign=c]{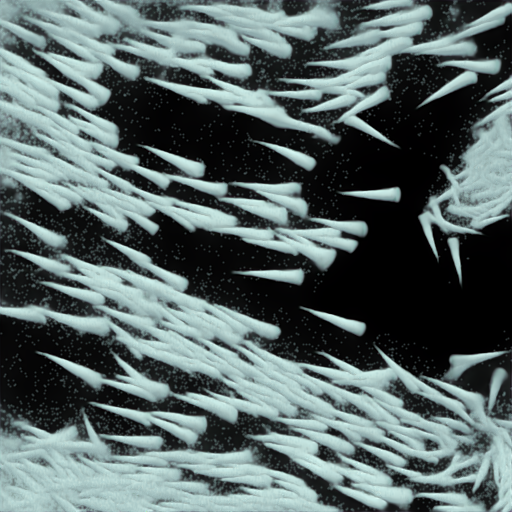} & \includegraphics[width=.38\linewidth,valign=c]{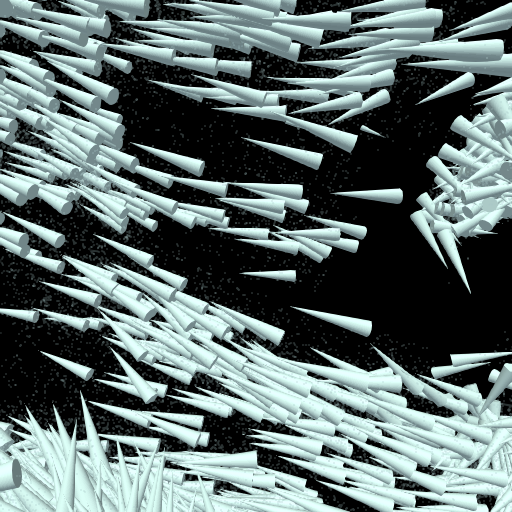}
  \vspace{3mm}\\
  \includegraphics[width=.38\linewidth,valign=c]{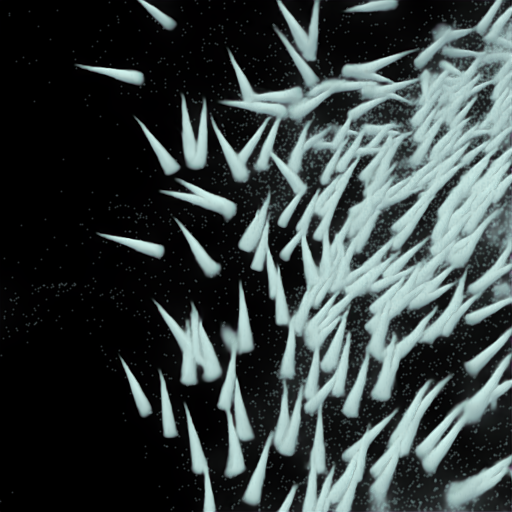} & \includegraphics[width=.38\linewidth,valign=c]{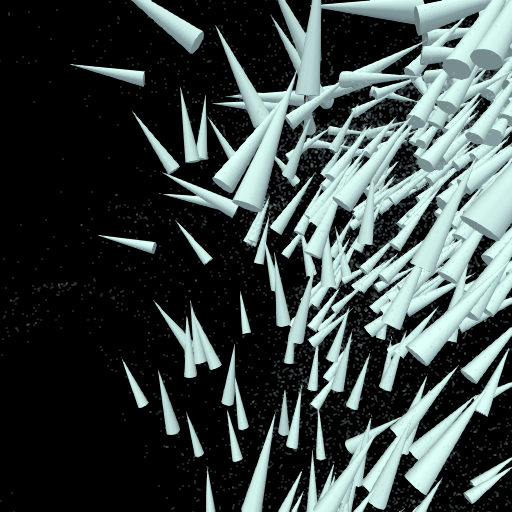}
  \vspace{3mm}\\
  \includegraphics[width=.38\linewidth,valign=c]{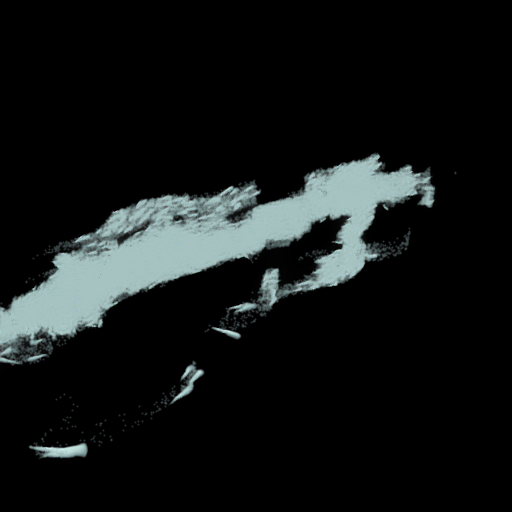} & \includegraphics[width=.38\linewidth,valign=c]{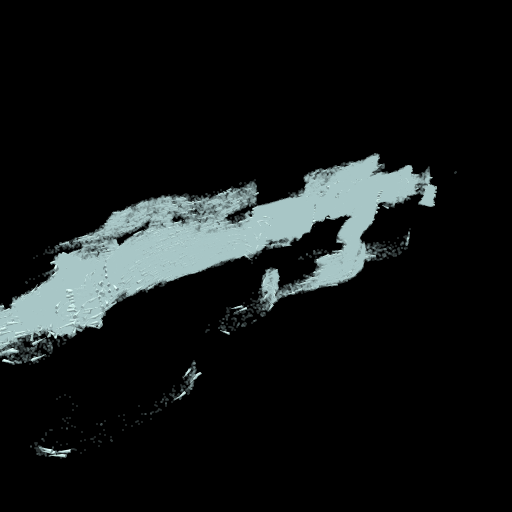}
  \vspace{3mm}\\
   (b) Target PC & (c) GT
  \end{tabular}
  \caption{Generalizability of NARVis Across Datasets. The post-processing effects of NARVis are transferrable across different datasets, assuming both the source and the target datasets have the same parameter mapping requirements for visualization. We learn from the reference Hurricane dataset (a) to visualize the Storms dataset (b). (c) Paraview ground truth rendering for reference.}
  \label{fig:genexp_glyph}
\end{figure}

We present the generalizability of NARVis when the source and the target PCs belong to different datasets. With a similar configuration as the partial data training case, we train NARVis on the source Hurricane dataset and transfer the visualization to the target Storms dataset. In this experiment, we visualize the velocity vectors in the Hurricane and Storms datasets with glyphs and without color mappings to ensure consistent visualization needs (see Figure \ref{fig:genexp_glyph}). We reduce the Storms resolution by $100\times$ for increased interpretability while visualizing without color mapping. We enable lighting and apply Gouraud shading on the glyphs to enhance the visualization's interpretability. We achieve an average PSNR of 17.819 and SSIM of 0.737 (w.r.t Paraview ground truth) for the target PC. We also observe from Figure \ref{fig:genexp_glyph} that NARVis enables transferring complex post-processing effects including glyphs, lighting, and shading effects while maintaining high visual fidelity.

\subsection{Ablation Studies}

\subsubsection{Ablation of the Framework Components}
\label{sec:abl_comp}
Table \ref{tab:perf} shows the breakdown of the runtimes of different components of our framework. \textit{Transfer + Proc.}  denotes the time required to convert OpenGL buffers to CUDA tensors, channel-wise processing, and downsampling the MACR output into 5 different resolutions. We observe that the U-Net inference is a major factor that causes latency. We implement the neural renderer in PyTorch~\cite{paszke2017automatic} without any neural acceleration. However, we observe a limited speedup and memory savings under certain cases on quantizing the U-Net (refer to the supplementary materials).
\\

\begin{table}[!h]
  \centering
  \begin{tabular}{%
  	  r |
  	  	*{4}{c}%
  	}
    \toprule
  	Component & \rotatebox{0}{Hurricane} & \rotatebox{0}{Storms} & Morro Bay\\
  	\midrule
        MACR & 0.104 & 0.106 & 0.122 \\
        Transfer + Proc. & 1.781 & 1.735 & 2.026 \\
        U-Net & 4.504 & 4.692 & 5.456 \\
    \bottomrule
    \end{tabular}
  \caption{%
  	Runtimes (in ms) of the different NARVis components evaluated at full PC resolution with $2000\times1328$ frame resolution.%
  }
  \label{tab:perf}
\end{table}

\noindent Using \texttt{$1\times1$-Conv} over using no descriptors and using per-point learnable neural descriptors (similar to \cite{aliev2020neural}) is beneficial (see Table \ref{tab:abunet}). Naively downsampling the feature images with neural descriptors results in large errors in the gradients. Thus, per-point descriptors require rendering the multi-scale feature images at 5 different resolutions resulting in a significant delay. Thus, using per-point descriptors slightly improves the rendering quality at the cost of $\sim4\times$ rendering delay. In Table \ref{tab:abunet}, columns \textit{PC} and \textit{Network} show the memory requirements for PC loading and NN parameters, respectively. \textit{PC} is calculated as $N \sum_i B_i$, where $N$ is the number of points, $B_i$ is the number of bytes for representing the $i^{th}$ data attribute. Both \texttt{$1\times1$-Conv} and \textit{No Descriptors} variants use point coordinates (3D float vector), byte-packed RGB, depth (single float) and velocity (4D float vector, see \textit{Vel2D} in section \ref{sec:data_streams}). \textit{Per-point Descriptor} variant represents each point with point coordinates (3D float vector), point indices (integers), and 4D learnable per-point descriptors (floats). \textit{Per-point Descriptors} variant, although with a marginally better rendering quality than \texttt{$1\times1$-Conv}, is unscalable as the memory requirements of the network depend on the number of points. \texttt{$1\times1$-Conv} improves the rendering quality and speed almost matching the memory overhead of the \textit{No Descriptors} case.
\\


\begin{table}[!h]
  \centering
  \begin{tabular}{%
  	  c |%
  	  	*{4}{c}%
  	}
    \toprule
    & \begin{tabular}{@{}c@{}}No\\Desc.\end{tabular} & \begin{tabular}{@{}c@{}}Per-point\\Desc.\end{tabular} & \begin{tabular}{@{}c@{}}$1 \times 1$\\Conv\end{tabular} \\
    \midrule
    PSNR & 22.24 & \textbf{22.56} & 22.42 \\
    SSIM & 0.761 & 0.812 & \textbf{0.814} \\
    Runtime (ms) & 7.077 & 24.88 & \textbf{6.971} \\
    PC Memory (MB) & 14.71 & \textbf{13.08} & 14.71 \\
    U-Net Memory (MB) & \textbf{7.73} & 14.35 & 7.74 \\
    \bottomrule
    \end{tabular}
  \caption{%
  	Ablation study of the U-Net variants evaluated on a full resolution Hurricane dataset with $2000\times1328$ frame resolution.%
  }
  \label{tab:abunet}
\end{table}

\noindent \textbf{Ablation of the loss function} Table \ref{tab:lossfn} shows how the different loss terms affect the rendering quality. We train NARVis on the full PC resolution of the Hurricane dataset. We observe that including (\checkmark) the $\mathcal{L}_{perc}$ term greatly improves the rendering fidelity, as it forces the network to learn visual features in the latent space~\cite{johnson2016perceptual}. Only using $\mathcal{L}_{reco}$ and $\mathcal{L}_{\nabla}$ is insufficient to achieve high rendering quality, as observed by low SSIM scores. Only using $\mathcal{L}_{\nabla}$ results in network learning scene textures only, resulting in inconsistent color distribution. Thus, for obtaining high-fidelity rendering we use all three loss terms.
\\

\begin{table}[!htb]
    \centering
    \begin{tabular}{%
          r%
              *{5}{c}%
        }
      \toprule
      $\mathcal{L}_{perc}$ & $\mathcal{L}_{reco}$ & $\mathcal{L}_{\nabla}$ & PSNR & SSIM \\
      \midrule
        \checkmark & \xmark & \xmark & 21.10 & 0.779 \\
        \xmark & \checkmark & \xmark & 21.40 & 0.718 \\
        \xmark & \xmark & \checkmark & 18.64 & 0.359 \\
        \checkmark & \checkmark & \xmark & 24.15 & 0.790 \\
        \checkmark & \xmark & \checkmark & 23.26 & 0.767 \\
        \xmark & \checkmark & \checkmark & 21.78 & 0.598 \\
        \checkmark & \checkmark & \checkmark & \textbf{24.65} & \textbf{0.813} \\
      \bottomrule
    \end{tabular}
    \caption{%
        Effect of different loss terms on the rendering performance on the Hurricane dataset at full PC resolution.%
    }
    \label{tab:lossfn}
\end{table}

\subsubsection{Effects of Point Cloud Resolution on Rendering Performance}
\noindent PC resolution is crucial for PC rendering. For example, a high-resolution PC can be rendered with high quality as it typically has denser regions with rich textures. However, rendering such high-resolution PCs is expensive and slow. NARVis allows using a down-sampled PC while maintaining the rendering quality and reducing memory footprints and rendering latency. We show this by varying the input PC resolution and recording the rendering performances. We reduce the PC resolution by uniformly sub-sampling the PC by different fixed factors. However, we still use $I_O$ generated with the full resolution PC while training.
\\

\begin{figure*}[!ht]
    \centering
    \begin{tabular}{cccccc}
    \vspace{3mm}
    \rotatebox[origin=c]{90}{Hurricane} & \includegraphics[width=.15\linewidth,valign=c]{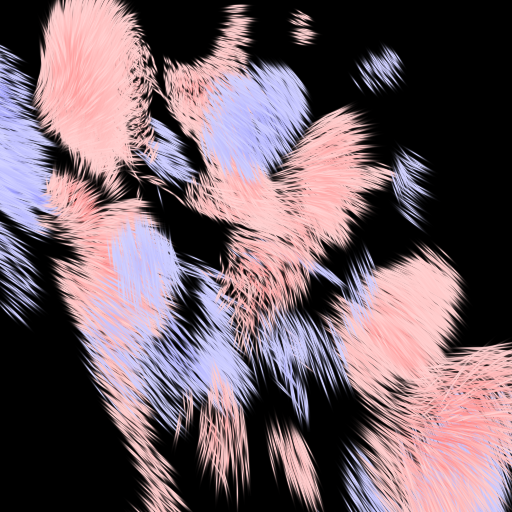} & \includegraphics[width=.15\linewidth,valign=c]{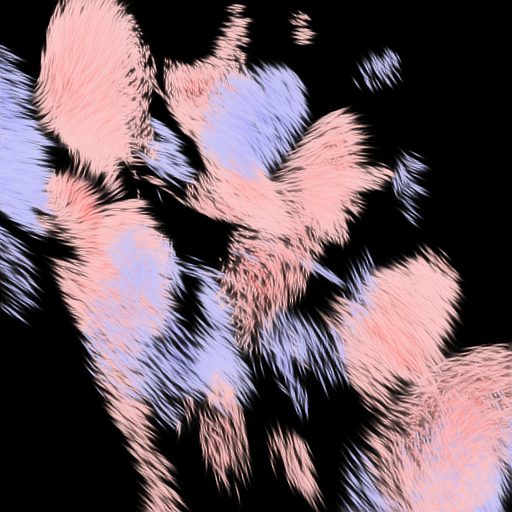} & \includegraphics[width=.15\linewidth,valign=c]{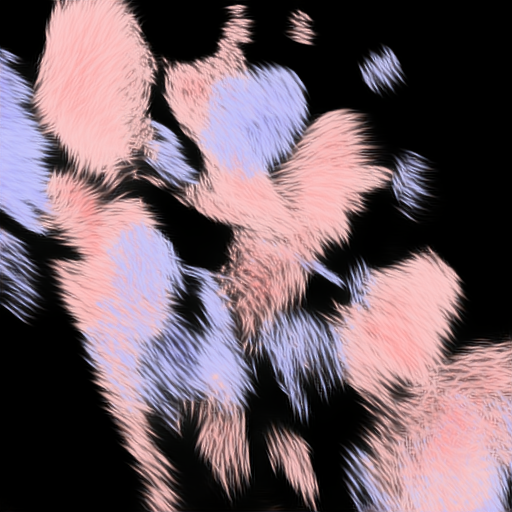} & \includegraphics[width=.15\linewidth,valign=c]{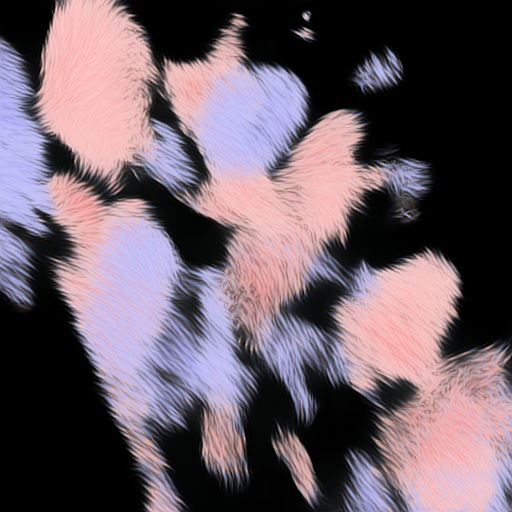} & 
    \includegraphics[width=.15\linewidth,valign=c]{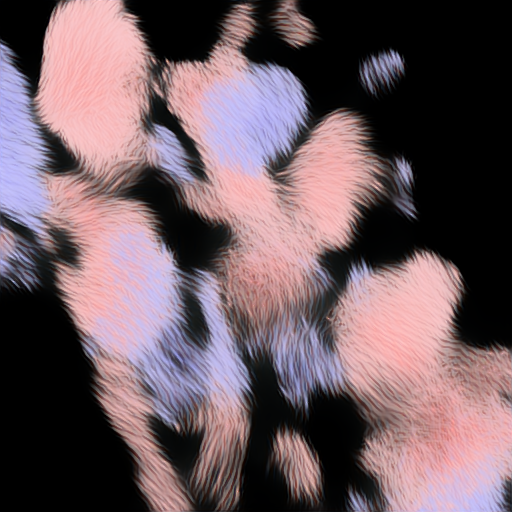}\\
    \vspace{3mm}
    \rotatebox[origin=c]{90}{Storms} & \includegraphics[width=.15\linewidth,valign=c]{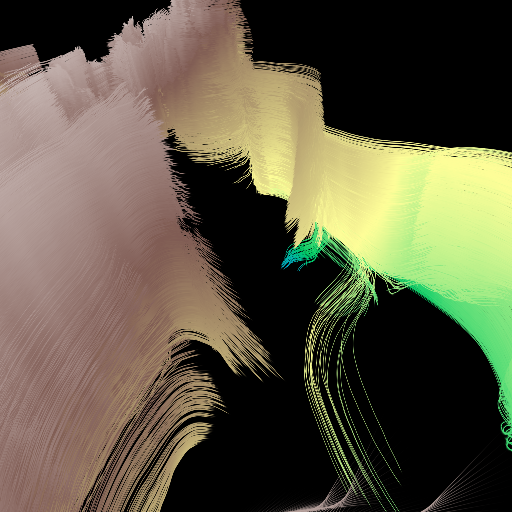} & \includegraphics[width=.15\linewidth,valign=c]{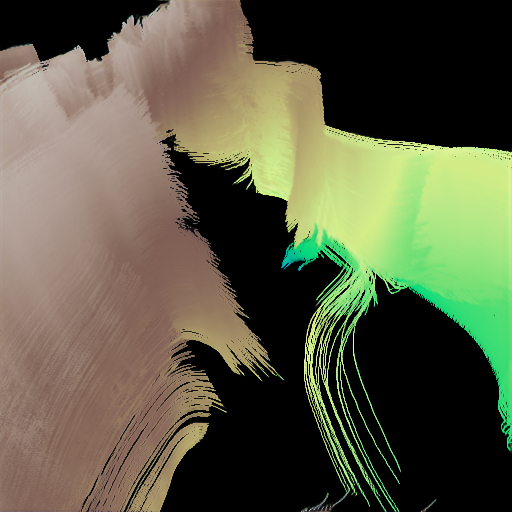} & \includegraphics[width=.15\linewidth,valign=c]{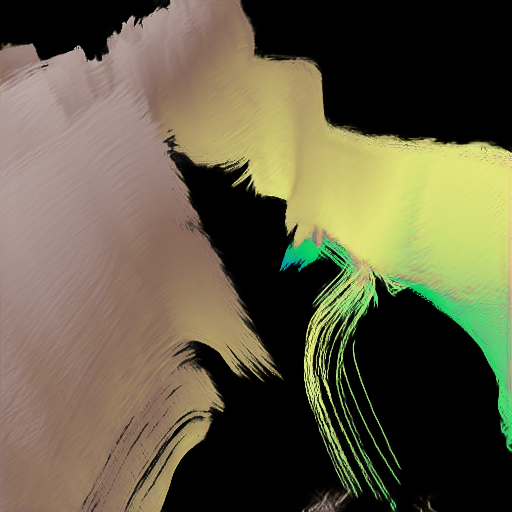} & \includegraphics[width=.15\linewidth,valign=c]{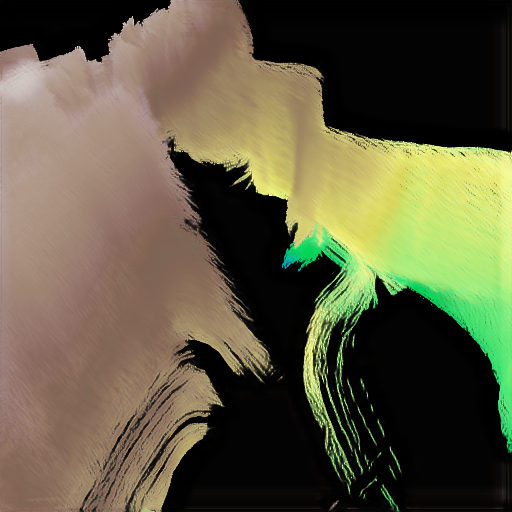} & \includegraphics[width=.15\linewidth,valign=c]{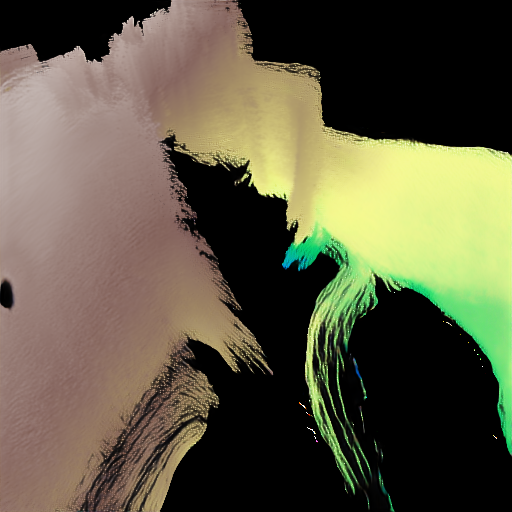}\\
    \vspace{3mm}
    \rotatebox[origin=c]{90}{Morro Bay} & \includegraphics[width=.15\linewidth,valign=c]{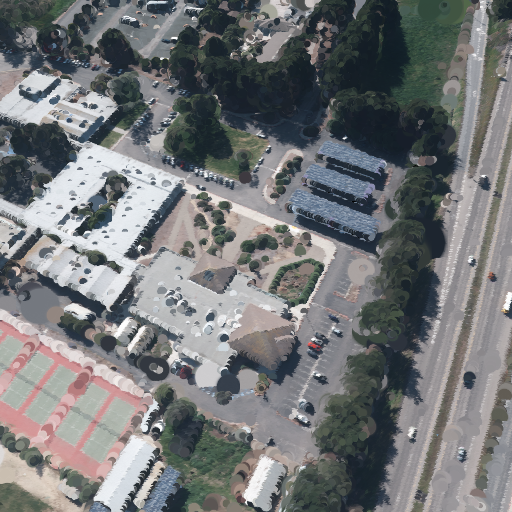} & \includegraphics[width=.15\linewidth,valign=c]{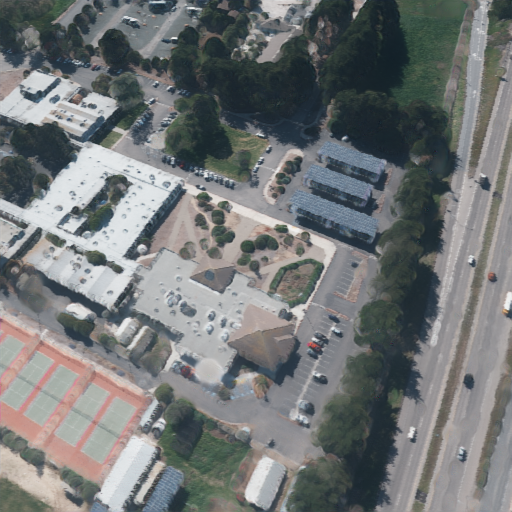} & \includegraphics[width=.15\linewidth,valign=c]{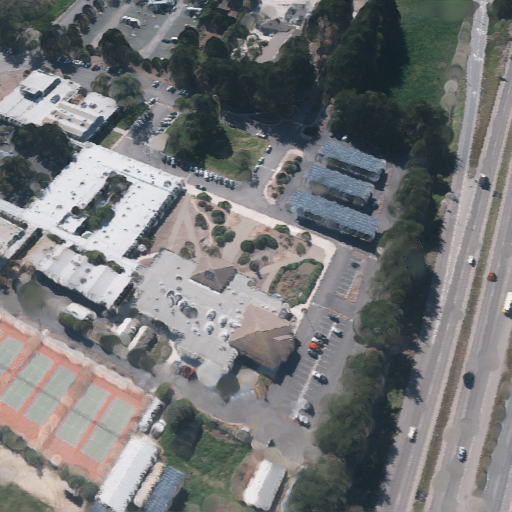} & \includegraphics[width=.15\linewidth,valign=c]{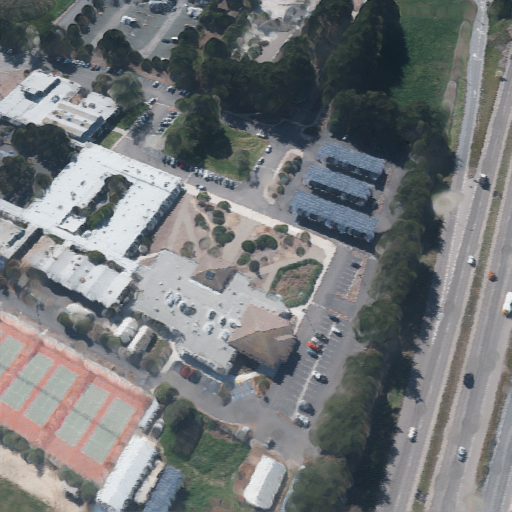} & \includegraphics[width=.15\linewidth,valign=c]{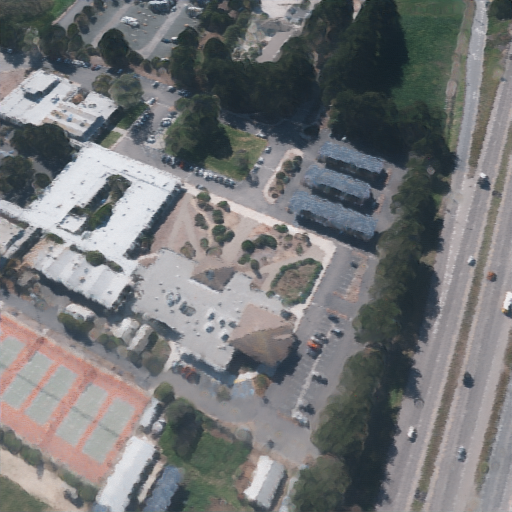}\\
    & (a) GR & (b)1x & (c) 2x & (d) 4x & (e) 10x
    \end{tabular}
    \caption{Qualitative results at various PC sparsity when compared with the GR ground truth. More results in the supplementary materials.}
    \label{fig:sr}
\end{figure*}

\noindent With decreasing PC resolution (or increased sparsity), rendering quality dips, but frame rates are nearly constant with reduced memory footprints (see Table \ref{tab:sr} and Figure \ref{fig:sr}). The dip in rendering quality is caused by the increased low-density regions in the PC, which are harder to interpolate for the neural renderer. We observe this in the drastic fall of the rendering qualitative metrics for a sparse Hurricane dataset. The neural renderer depends on the number of points in the projected frame and not on the overall points in the PC, explaining the near-constant frame rates, irrespective of the dataset used. However, for larger datasets, such as Morrobay, the frame rate increase is pronounced with a reduction in PC resolution. Furthermore, due to a significant fixed G-buffer overhead, we do not see a proportional decrease of memory consumption. 

\begin{table}[!h]
    \centering
    \begin{tabular}{%
          c%
              *{6}{c}%
        }
    \toprule
        Dataset & \rotatebox{0}{S} & \rotatebox{0}{PSNR} &   \rotatebox{0}{SSIM} &   \makecell{Latency \\(ms) / fps} & \makecell{Memory \\(GB)}  \\
        \midrule\
          \multirow{4}{*}{\rotatebox{90}{\makecell{\textit{Hurricane} \\ \textit{(408K pts)}}}}{} & 1x & \textbf{24.65} & \textbf{0.813}  &  6.97 / 143 & 1.7 \\
          & 2x & 22.61 & 0.743  &  \textbf{6.77 / 147} & 1.6 \\
          & 4x & 21.19 & 0.701  &  6.79 / 147 & 1.6 \\
          & 10x & 20.21 & 0.673 &  6.89 / 145 & 1.6 \\
          \midrule
          \multirow{4}{*}{\rotatebox{90}{\makecell{\textit{Storms} \\ \textit{(45M pts)}}}} & 1x &  \textbf{28.02} & \textbf{0.915}  &  \textbf{6.58 / 152} & 3.1 \\
          & 2x &  24.29 & 0.840  &  6.62 / 151 & 2.4 \\
          & 4x &  23.62 & 0.792  &  7.07 / 141 & 2.1 \\
          & 10x &  23.40 & 0.706  &  6.79 / 147 & 1.9 \\
          \midrule
          \multirow{4}{*}{\rotatebox{90}{\makecell{\textit{Morro Bay} \\ \textit{(350M pts)}}}} & 1x &  \textbf{24.49} & 0.683  & 7.91 / 126 & 12.7 \\
          & 2x &  24.09 & \textbf{0.686}  &  7.36 / 136 & 9.4 \\
          & 4x &  23.73 & 0.641 &  7.18 / 139 & 7.8 \\
          & 10x &  23.18 & 0.608 &  \textbf{7.15 / 140} & 6.7 \\
    \bottomrule
    \end{tabular}
    \caption{%
        Evaluating PC sub-sampling effects on performances of the proposed method on different datasets. (\textit{S} is the PC sparsity.)%
    }
    \label{tab:sr}
\end{table}

\subsubsection{Influence of Data Attributes on Rendering Quality}
\label{sec:data_streams}

In Table \ref{tab:streams}, we pass combinations of different attributes and study their effects on neural rendering quality. To demonstrate the effects of different attributes, we specifically visualize the velocity vector, $\mathbf{v}$, associated with each point. Here, \textit{RGB} are the color channels depicting the photometric properties of the data, \textit{D} is the scene depth w.r.t the camera, \textit{Vel3D} represents a 4D vector $\mathbf{v_{3D}} = (v_x, v_y, v_z, |\mathbf{v}|)$ where $v_x, v_y, v_z$ are the 3D velocity components, and \textit{Vel2D} is another 4D vector $\mathbf{v_{2D}} = (v'_x, v'_y, \theta, |\mathbf{v'}|)$, where $\mathbf{v'}$ is the perspective projected $\mathbf{v}$ vector and $v'_x, v'_y$ are the 2D components of $\mathbf{v'}$, and $\theta$ is the angle of inclination of $\mathbf{v'}$ in the image plane. We observe a marginal improvement in rendering quality by adding depth information. Passing \textit{Vel2D} to the U-Net performs the best as \textit{Vel2D} explicitly encodes the geometric cues of the point. \textit{Vel3D} does not improve the rendering quality as it is view-independent information and, thus, is less informative than the perspective projected \textit{Vel2D}. Also refer to Figure \ref{fig:streams} in the supplementary materials for qualitative results of the different data attributes. Baseline geometry and depth of the PC provide the majority of the reconstruction signal. Attributes like velocity contribute semantic refinement rather than structural changes. Thus, we observe similar quantitative performances of all these variants.

\begin{table}[!htb]
    \centering
    \begin{tabular}{%
          r%
              *{3}{c}%
        }
      \toprule
      Attributes & PSNR & SSIM \\
      \midrule
        RGB & 23.03 & 0.732 \\
        RGB+D & 23.19 & 0.741 \\
        RGB+Vel2D & \textbf{25.13} & 0.812 \\
        RGB+D+Vel3D & 22.92 & 0.754 \\
        RGB+D+Vel2D & 24.65 & \textbf{0.813} \\
      \bottomrule
    \end{tabular}
    \caption{%
        Evaluating our framework with different input data attributes of the Hurricane dataset at full PC resolution.%
    }
    \label{tab:streams}
\end{table}

\subsection{Limitations}
While NARVis enables real-time visualization of massive point clouds, it has several limitations. First, the MACR currently relies on early z-culling for efficiency, and, therefore, it does not handle transparency effects explicitly. However, transparency is implicitly handled by the post-processing network, which could be less reliable. Second, the U-Net struggles to reconstruct intricately detailed glyph varieties with non-convex shapes or extremely sharp high-frequency features such as arrows. Third, we may observe minor temporal flickering in the interactive viewing of certain datasets because NARVis processes frames independently without explicitly enforcing temporal consistency. Finally, in cases of varied parameter mappings (for example, Gaussians used differently in Hurricane and Morro Bay datasets), we must retrain NARVis for each visualization separately, independent of the PC geometry. Therefore, for practical purposes, we can collate a model zoo trained on different parameter mappings that could be loaded at runtime.


\subsection{Discussion}

To the best of our knowledge, no previous work addresses PC post-processing to accommodate different visualization needs, especially for scientific PCs. The post-processing requirements could be controlled using different conventional renderers of the users' choice (Paraview~\cite{ayachit2015paraview} and GR - 3DGS's~\cite{kerbl3Dgaussians} forward renderer - in our case). Furthermore, the choice of the specific data attributes and their mode of representation (i.e, point attributes such as color, position, glyph, Gaussian anisotropy, etc.) for visualizing a particular dataset rests with the user. NARVis, regardless of attribute requirements for visualization, accelerates the rendering process compared to the reference conventional renderer and is scalable to large PCs (with >350M points). Under the same parameter mappings, NARVis is also generalizable across different PC datasets and geometries (section \ref{sec:general}).
\\

\noindent General volume rendering formulations, such as volume ray marching~\cite{levoy2002display}, Gaussian splatting~\cite{kerbl3Dgaussians}, maximum intensity projection~\cite{wallis1989three}, etc., are high-quality volume rendering techniques that diverge from the PC visualization/rendering task. Thus, we avoid evaluating our methods against these formulations. We would also like to add a note on the GR's rendering performance. The rendering speed of the GR is sensitive to the camera view as it controls the number of Gaussians present in the scene. For instance, in the Storms dataset, we observed a rendering time of 8ms (on average) for close-up views with less Gaussians ($\sim$10K). However, when we zoom out to include all the Gaussians ($\sim$45M) it can take up to 3.5s (on average) to render the image. Thus, for comparing GR's rendering speeds with other methods,we use standardized camera views of the renders to ensure fair evaluation.
\\

\noindent 


%% file: sections/conclusions.tex
\section{Conclusions}
We present NARVis, an efficient solution for the real-time high-fidelity visualization of massive scientific PCs. NARVis is ideal for diverse scientific visualizations such as complex Lagrangian dynamics and a massive photometric terrain scan. We demonstrate the ability to render 350M points (12.7GB) at $>$126 fps (i.e., a throughput of over 44B points/second) on an RTX 2080 Ti GPU. NARVis is generalizable for different PCs and retains substantial rendering quality even when used with a subsampled PC, ultimately reducing memory requirements. We evaluate and discuss the different optimization choices through an ablation study to significantly reduce rendering latency and memory footprints. NARVis's low motion-to-photon latency and high fidelity make it foundational for immersive visualizations to help scientists analyze scientific phenomena.

%% file: sections/appendix.tex
\newpage
\setcounter{page}{1}

\section*{Supplementary Material}

\section{Selecting Viewpoints in Training Data}
\label{app:viewpts}
\begin{figure}[!ht]
    \centering
    \subfigure[Viewpoints from Grid]{\includegraphics[width=0.7\linewidth]{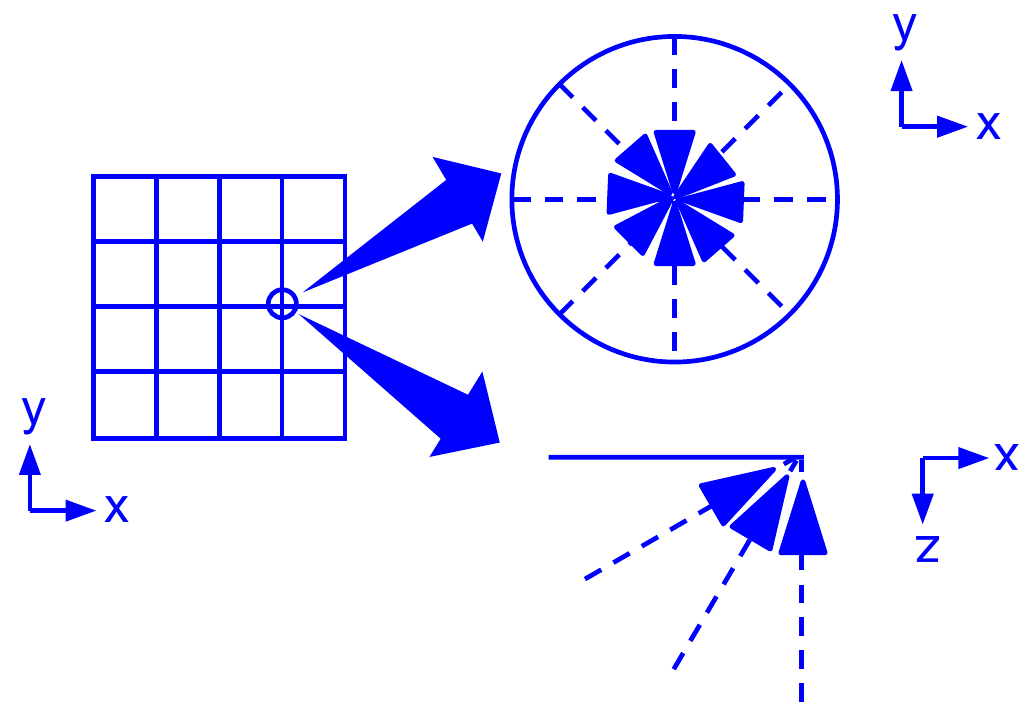}} 
    \subfigure[Viewpoints from Hemisphere]{\includegraphics[width=0.7\linewidth]{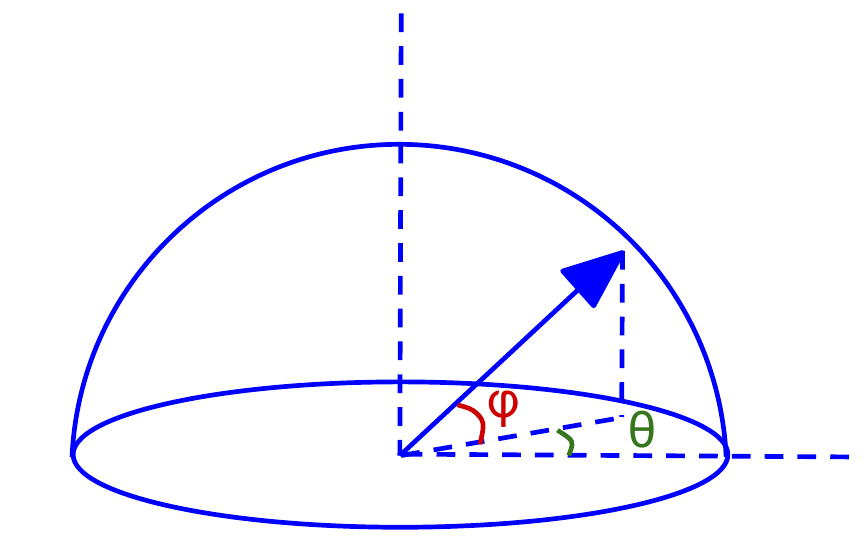}} 
    \caption{Strategies for Selecting Viewpoints for Training. Triangles in the diagram represent cameras with look-at vectors along the line passing through the base of the triangle.}
    \label{fig:dc_strat}
\end{figure}

\noindent Figure \ref{fig:dc_strat} shows the viewpoint sampling strategies for collecting the training and evaluation data at a $512 \times 512$ resolution. For all datasets, we select the viewpoints with two strategies. In the first strategy, we place our virtual cameras at the intersections of the $10 \times 10$ grid ($15 \times 15$ for Hurricane) which is parallel to the x,y-plane of the PC bounding box at a constant predetermined height above the maximum z coordinate of the bounding box. At each grid location, we capture the views by rotating the camera about the z-axis at the interval of $45^{\circ}$ (yaw) and also at the pitch levels $1^{\circ}, 30^{\circ},$ and $60^{\circ}$. Optionally, we sample multiple such grids at different heights above x,y-plane for including multi-scale details of the PC. In the second strategy, we consider a hemisphere around the PC. The hemisphere is centered at the center of the PC bounding box and the radius of the hemisphere, $r$, is $0.75\times$ of the PC bounding box diagonal. We then place the cameras on the hemisphere so that the polar coordinates of the cameras could be represented as $(r, \theta, \phi)$. Here, $\theta \in [0, 2\pi]$ sampled in the interval of $\pi/8$ ($\pi/16$ for Hurricane) and $\phi \in [0, \pi/2]$ sampled in the interval of $\pi/16$ ($\pi/32$ for Hurricane). Also, note that our selection strategy is not rigorous in coverage. For a formal study on view selection strategies, refer to ~\cite{xiao2024nerf}.


\section{U-Net Quantization}
\label{app:quant}
We report the results (see Table \ref{tab:appq}) on applying a static quantization post-training on the U-Net weights and activations by reducing the precision from \texttt{fp32} to \texttt{fp16}. This reduces the memory requirements by a factor of ~$1.18\times$ (on average across datasets) at the cost of a dip in rendering quality due to loss of precision. However, quantization marginally improves the inference speed only for larger PCs (see \textit{Morro Bay}, Table \ref{tab:appq}) which saturate the GPU memory.

\begin{table}[!ht]
  \centering
  \begin{tabular}{%
  	  r%
  	  	*{5}{c}%
  	}
    \toprule
        Dataset & \rotatebox{0}{PSNR} &   \rotatebox{0}{SSIM} & \makecell{Latency \\(in ms) / fps} & \makecell{Memory \\ (in GB)} \\
        \midrule
        Hurricane & 24.62 & 0.833 & 7.97 / 125.48 & 1.40 \\
        Storms & 19.73 & 0.808 & 7.88 / 126.90 & 2.80 \\
        Morro Bay & 22.27 & 0.654 & 7.82 / 127.81 & 10.40 \\
        
  \bottomrule
  \end{tabular}
  \caption{%
  	Comparison of rendering quality (w.r.t GSplat rendering), latency, and memory footprints of NARVis under quantization.%
  }
  \label{tab:appq}
\end{table}

\section{Additional Results}
\label{app:addres}

\begin{figure}[!h]
    \centering
    \begin{tabular}{cccc}
    \includegraphics[width=.45\linewidth,valign=c]{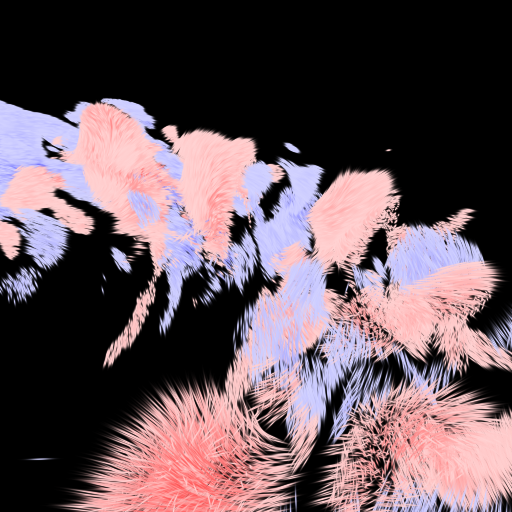} & \includegraphics[width=.45\linewidth,valign=c]{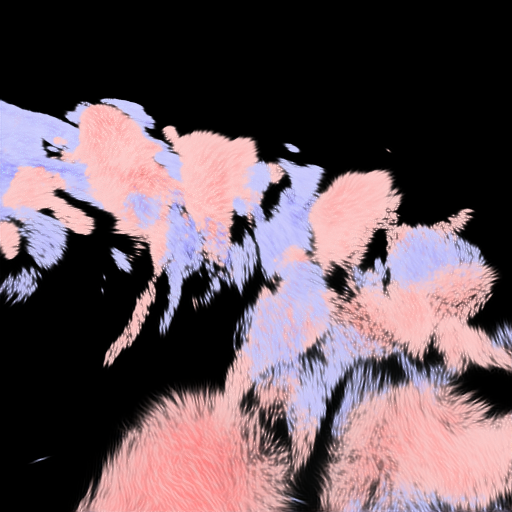}
    \vspace{2mm}\\
    (a) GR & (b) RGB
    \vspace{2mm}\\
    \includegraphics[width=.45\linewidth,valign=c]{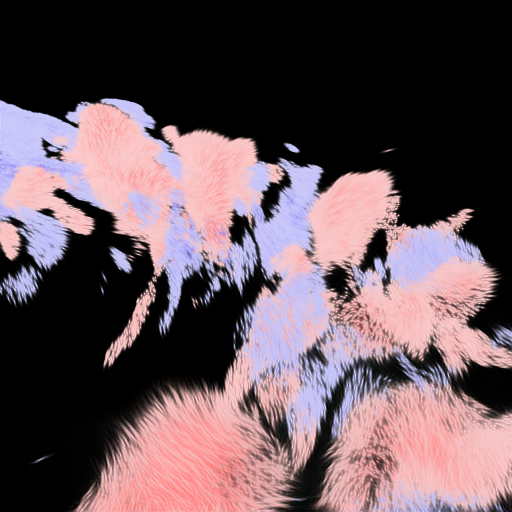} &
    \includegraphics[width=.45\linewidth,valign=c]{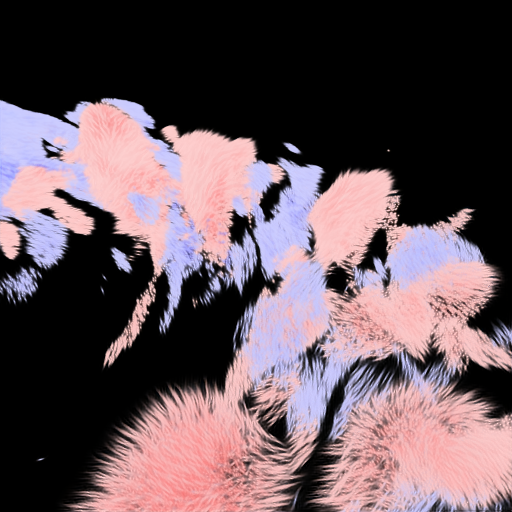}
    \vspace{2mm}\\
    (c) RGB+D & (d) RGB+Vel2D
    \vspace{2mm}\\
    \includegraphics[width=.45\linewidth,valign=c]{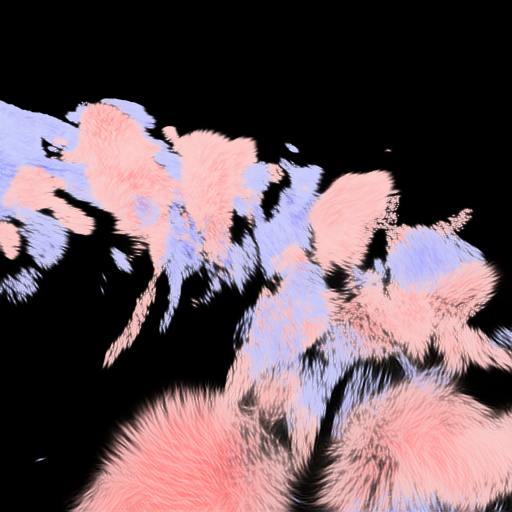} & \includegraphics[width=.45\linewidth,valign=c]{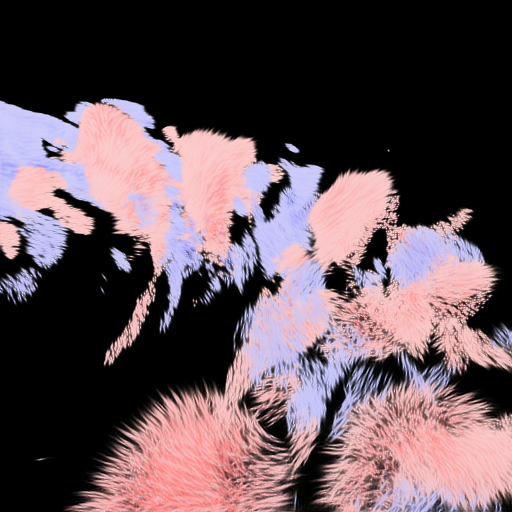}
    \vspace{2mm}\\
    (e) RGB+D+Vel3D & (f) RGB+D+Vel2D
    \end{tabular}
    \caption{Qualitative results of NARVis when UNet is trained with different input attributes from MACR. See Section \ref{sec:data_streams} for more details on the point attributes and their definition.}
    \label{fig:streams}
\end{figure}

\begin{figure*}[!ht]
    \centering
    \begin{tabular}{ccccc}
        \vspace{3mm}
        \rotatebox[origin=c]{90}{Hurricane} & \raisebox{-0.5\height}{\includegraphics[width=.18\linewidth]{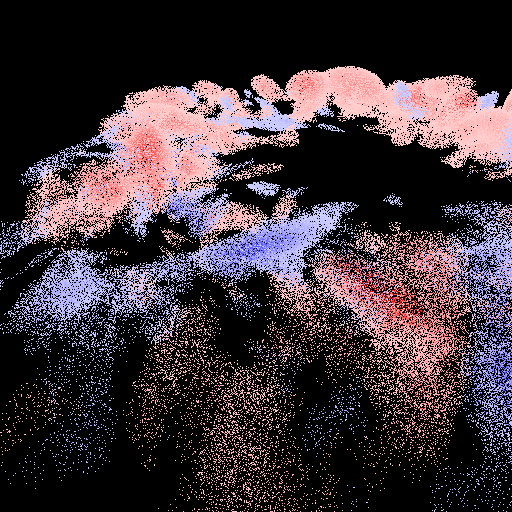}} & \raisebox{-0.5\height}{\includegraphics[width=.18\linewidth]{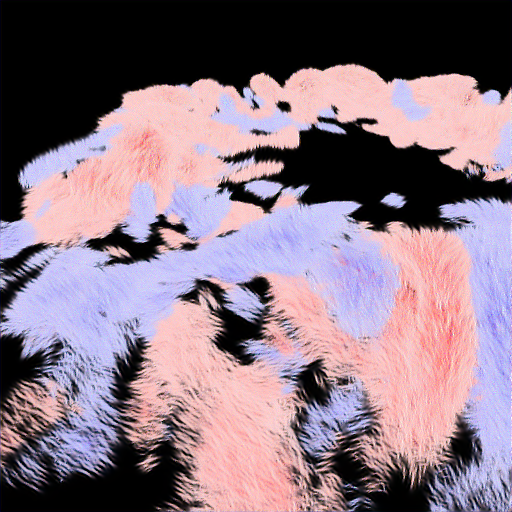}} & \raisebox{-0.5\height}{\includegraphics[width=.18\linewidth]{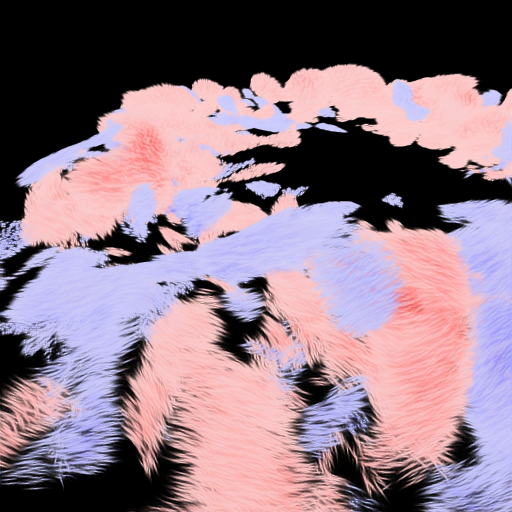}} & \raisebox{-0.5\height}{\includegraphics[width=.18\linewidth]{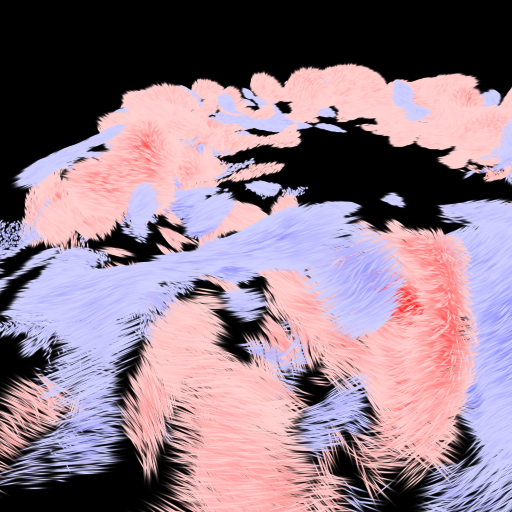}}\\
        \vspace{3mm}
        \rotatebox[origin=c]{90}{Hurricane} & \raisebox{-0.5\height}{\includegraphics[width=.18\linewidth]{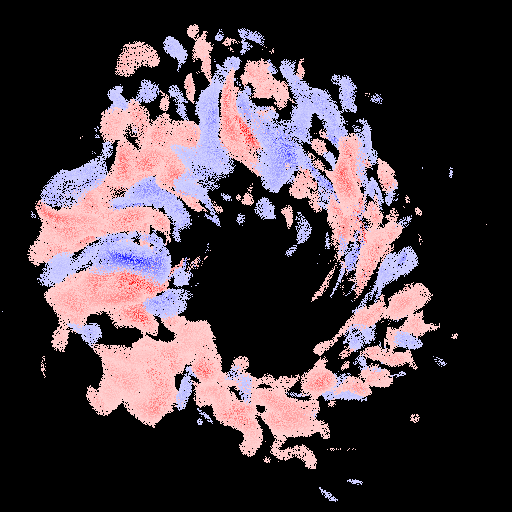}} & \raisebox{-0.5\height}{\includegraphics[width=.18\linewidth]{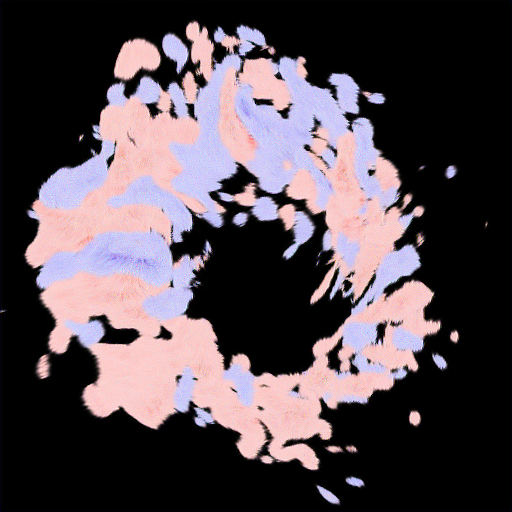}} & \raisebox{-0.5\height}{\includegraphics[width=.18\linewidth]{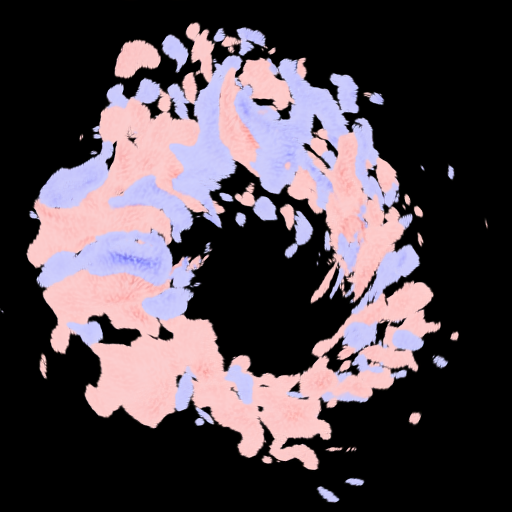}} & \raisebox{-0.5\height}{\includegraphics[width=.18\linewidth]{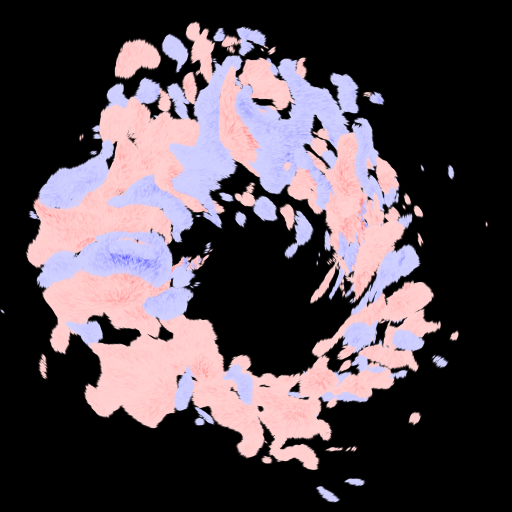}}\\
        \vspace{3mm}
        \rotatebox[origin=c]{90}{Storms} & \raisebox{-0.5\height}{\includegraphics[width=.18\linewidth]{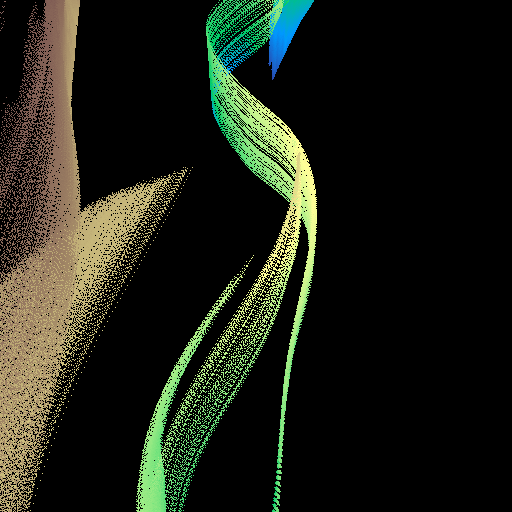}} & \raisebox{-0.5\height}{\includegraphics[width=.18\linewidth]{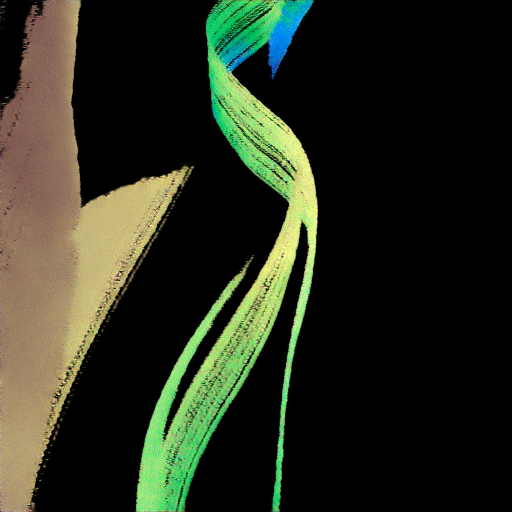}} & \raisebox{-0.5\height}{\includegraphics[width=.18\linewidth]{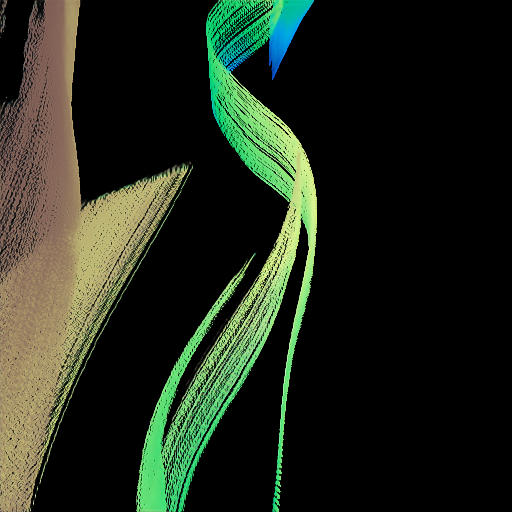}} & \raisebox{-0.5\height}{\includegraphics[width=.18\linewidth]{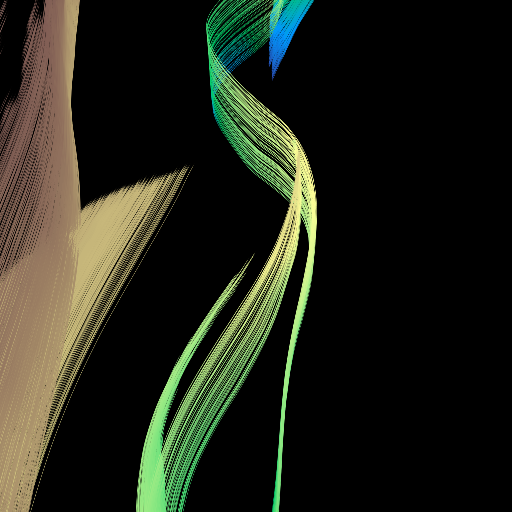}}\\
        \vspace{3mm}
        \rotatebox[origin=c]{90}{Storms} & \raisebox{-0.5\height}{\includegraphics[width=.18\linewidth]{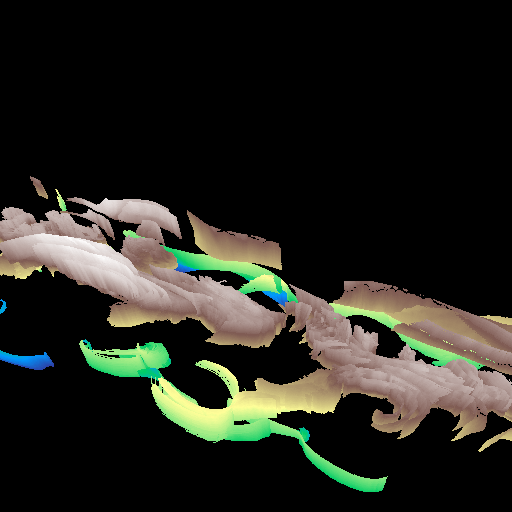}} & \raisebox{-0.5\height}{\includegraphics[width=.18\linewidth]{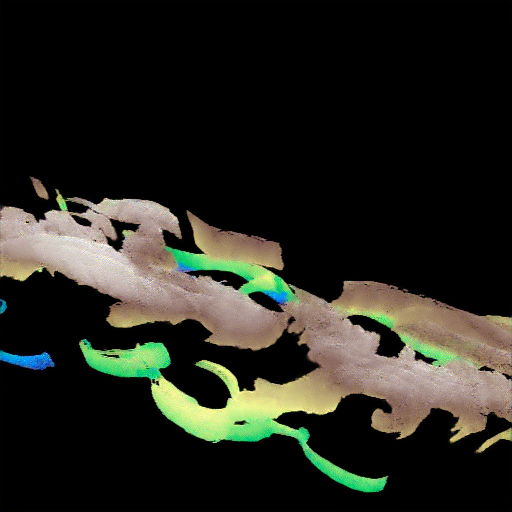}} & \raisebox{-0.5\height}{\includegraphics[width=.18\linewidth]{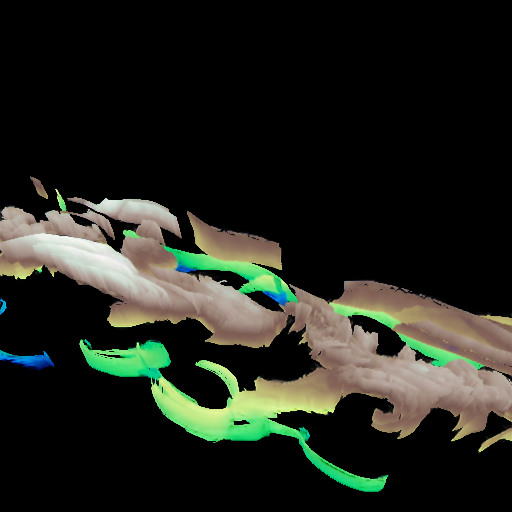}} & \raisebox{-0.5\height}{\includegraphics[width=.18\linewidth]{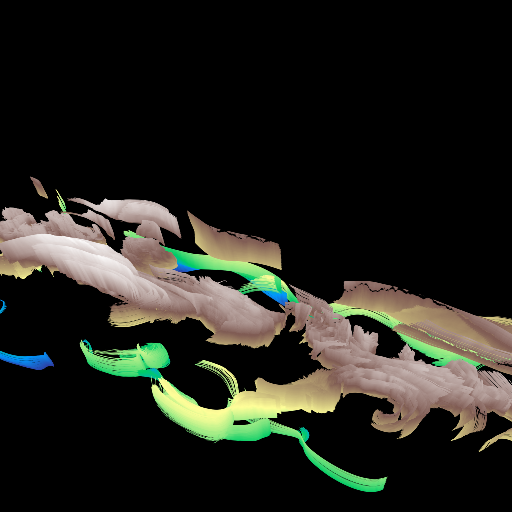}}\\
        \vspace{3mm}
        \rotatebox[origin=c]{90}{Morro Bay$^\dagger$} & \raisebox{-0.5\height}{\includegraphics[width=.18\linewidth]{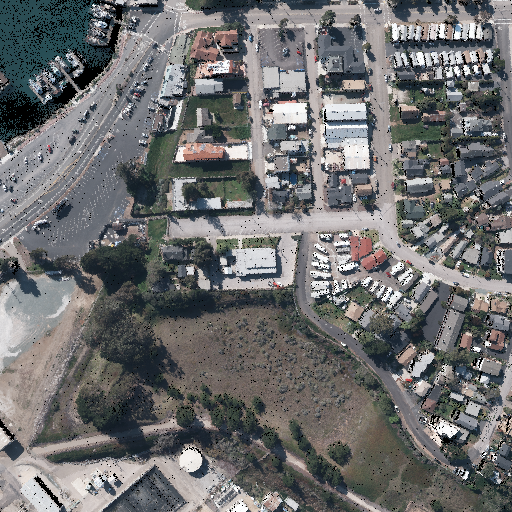}} & \raisebox{-0.5\height}{\includegraphics[width=.18\linewidth]{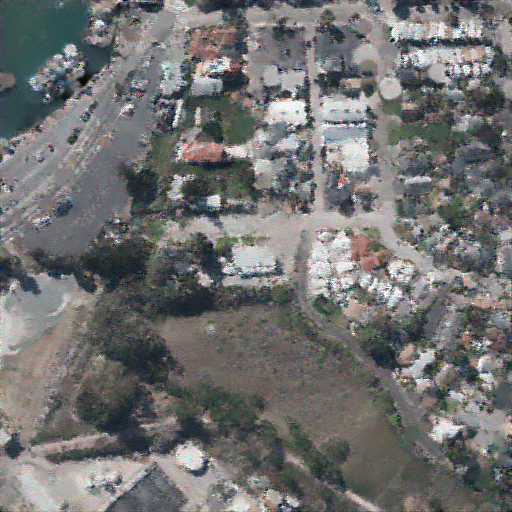}} & \raisebox{-0.5\height}{\includegraphics[width=.18\linewidth]{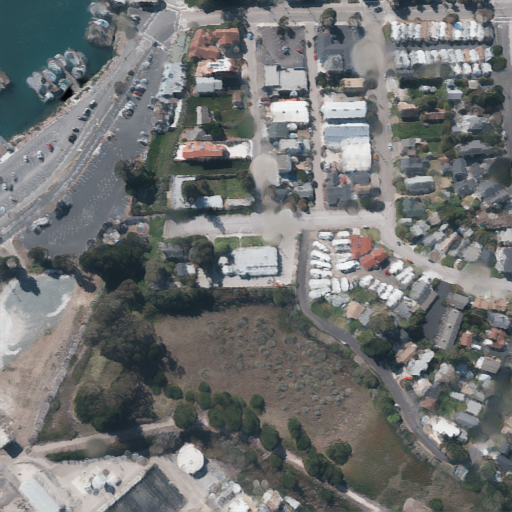}} & \raisebox{-0.5\height}{\includegraphics[width=.18\linewidth]{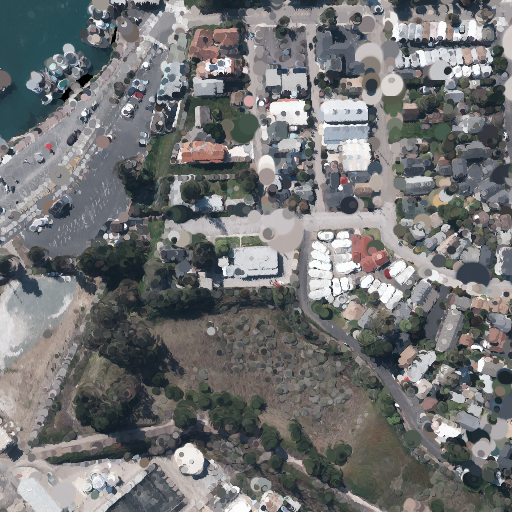}}\\
        \vspace{3mm}
        \rotatebox[origin=c]{90}{Morro Bay$^\dagger$} & \raisebox{-0.5\height}{\includegraphics[width=.18\linewidth]{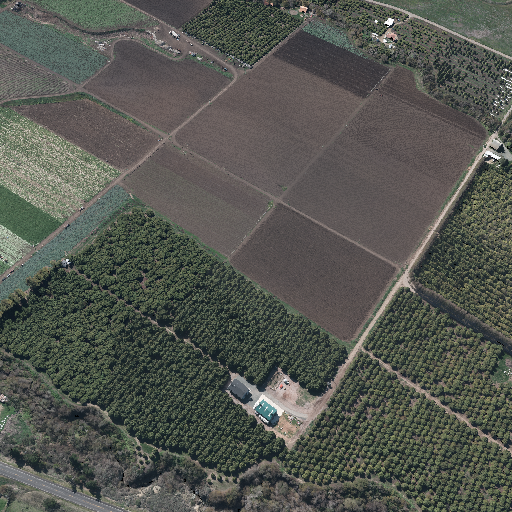}} & \raisebox{-0.5\height}{\includegraphics[width=.18\linewidth]{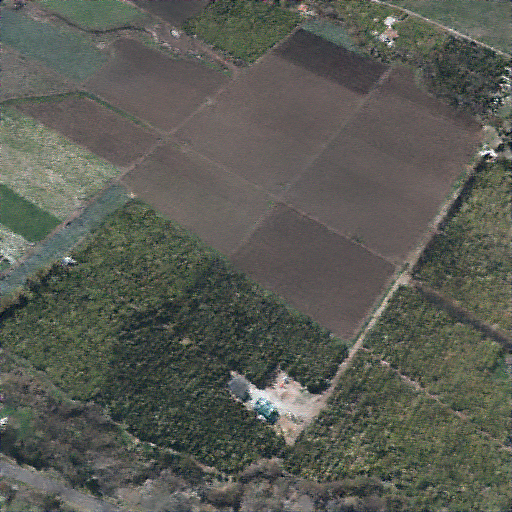}} & \raisebox{-0.5\height}{\includegraphics[width=.18\linewidth]{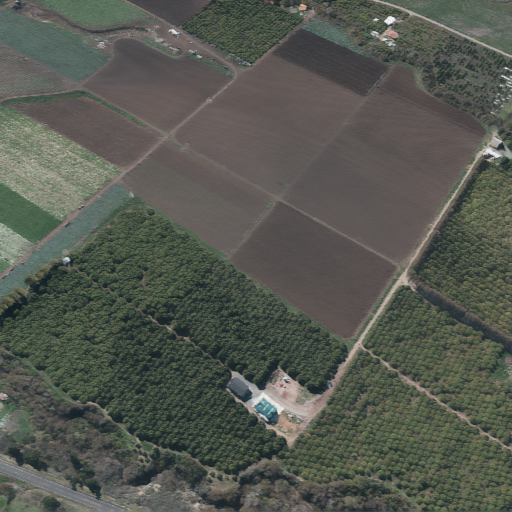}} & \raisebox{-0.5\height}{\includegraphics[width=.18\linewidth]{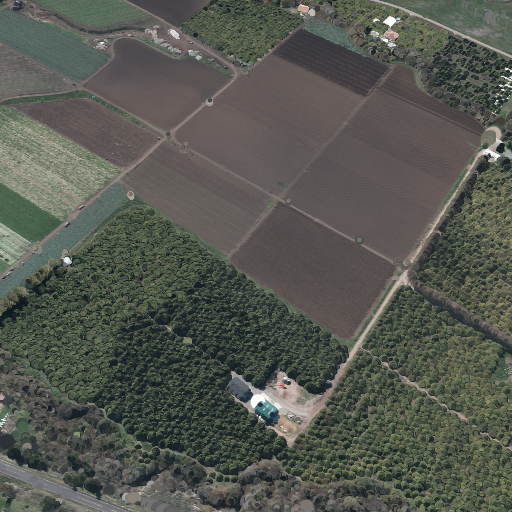}}\\
        & (a) CR & (b) NPBG & (c) Ours & (d) GSplat
    \end{tabular}
    \caption{Qualitative comparison of different rendering methods. Both NPBG and NARVis are trained on GSplat rendered images. $^\dagger$For the MorroBay dataset, GSplat and NPBG used 0.5× and 0.1× the original number of points, respectively, due to memory restrictions.}
    \label{fig:app_comp}
\end{figure*}

\begin{figure*}[!ht]
    \centering
    \begin{tabular}{cccccc}
    \vspace{3mm} 
    \rotatebox[origin=c]{90}{Hurricane} & \includegraphics[width=.17\linewidth,valign=c]{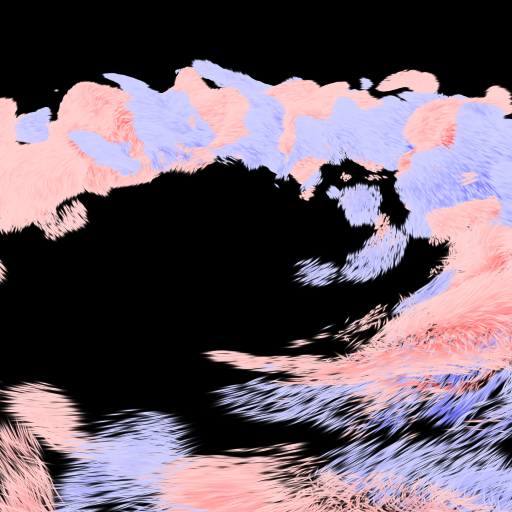} & \includegraphics[width=.17\linewidth,valign=c]{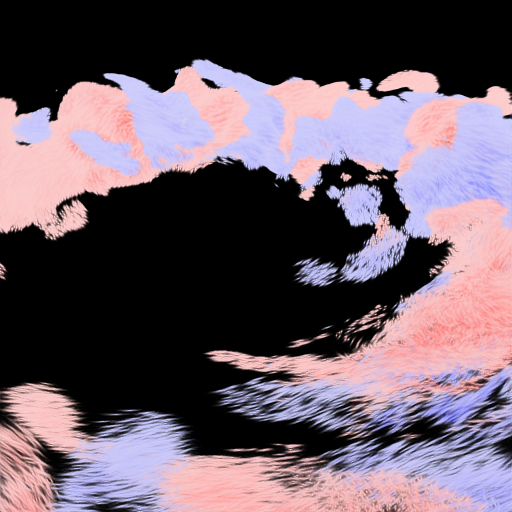} & \includegraphics[width=.17\linewidth,valign=c]{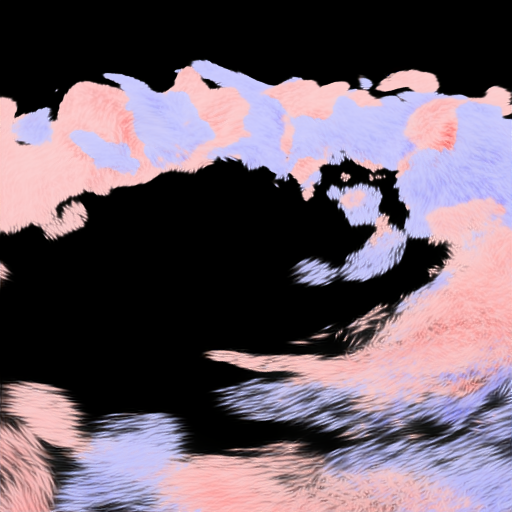} & \includegraphics[width=.17\linewidth,valign=c]{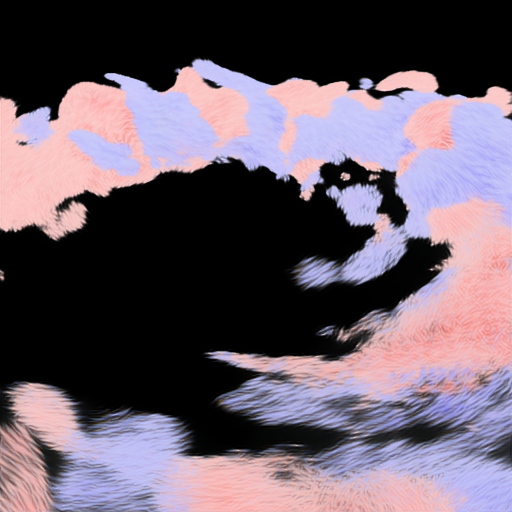} & 
    \includegraphics[width=.17\linewidth,valign=c]{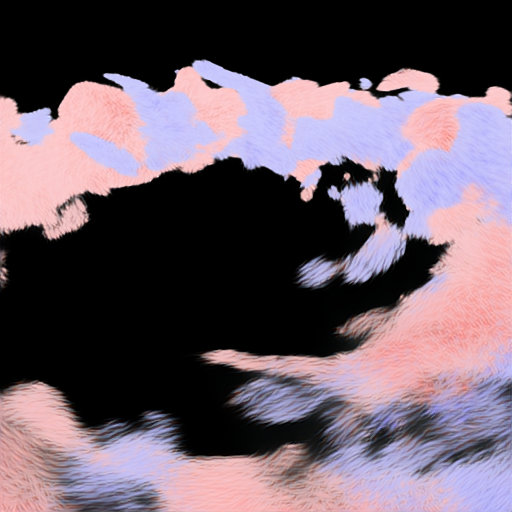}\\
    \vspace{3mm} 
    \rotatebox[origin=c]{90}{Hurricane} & \includegraphics[width=.17\linewidth,valign=c]{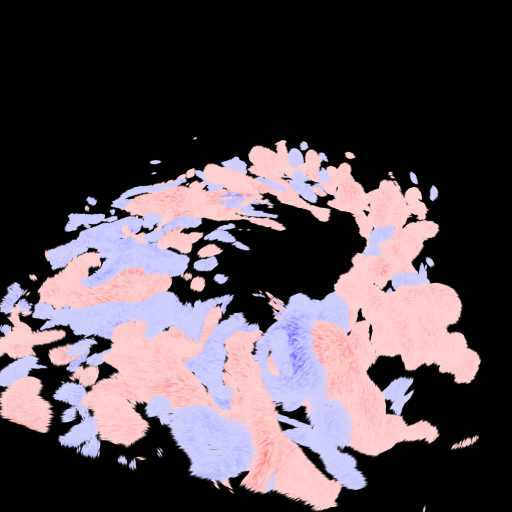} & \includegraphics[width=.17\linewidth,valign=c]{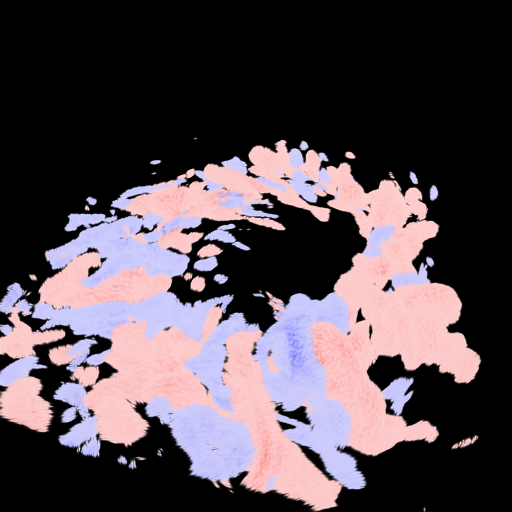} & \includegraphics[width=.17\linewidth,valign=c]{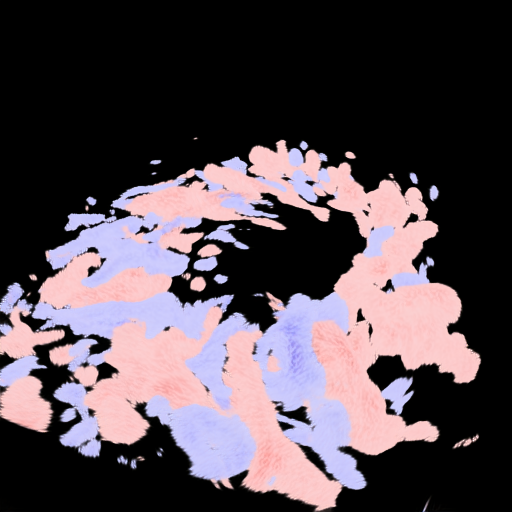} & \includegraphics[width=.17\linewidth,valign=c]{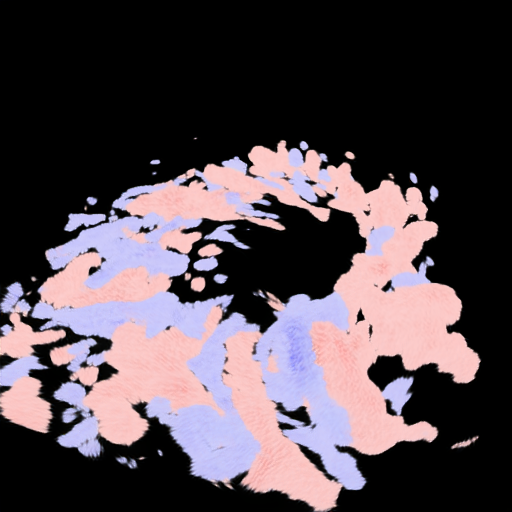} & 
    \includegraphics[width=.17\linewidth,valign=c]{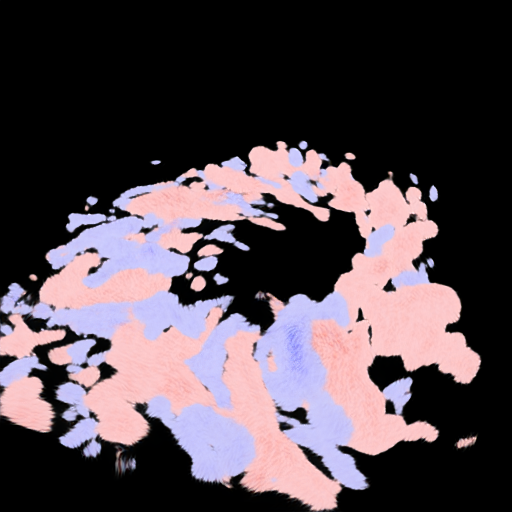}\\
    \vspace{3mm} 
    \rotatebox[origin=c]{90}{Storms} & \includegraphics[width=.17\linewidth,valign=c]{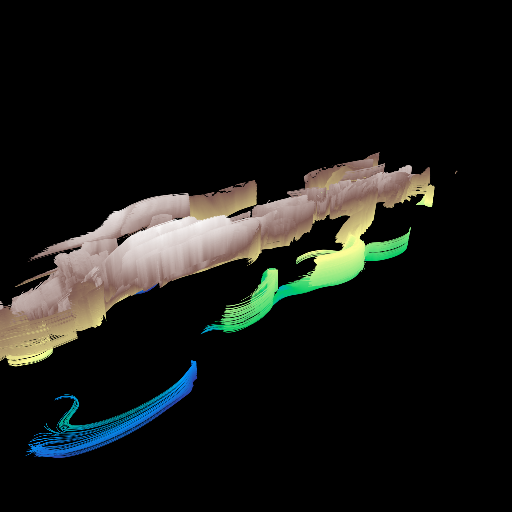} & \includegraphics[width=.17\linewidth,valign=c]{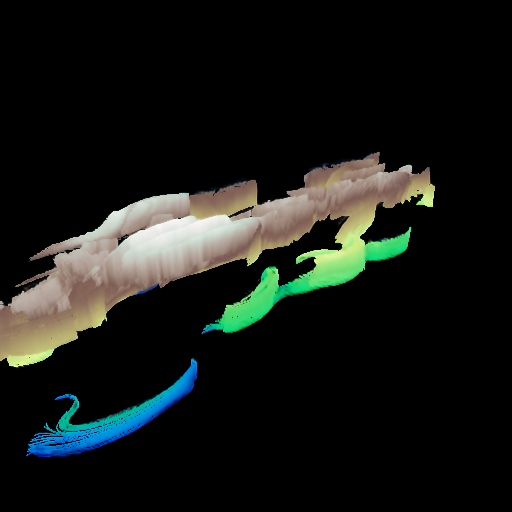} & \includegraphics[width=.17\linewidth,valign=c]{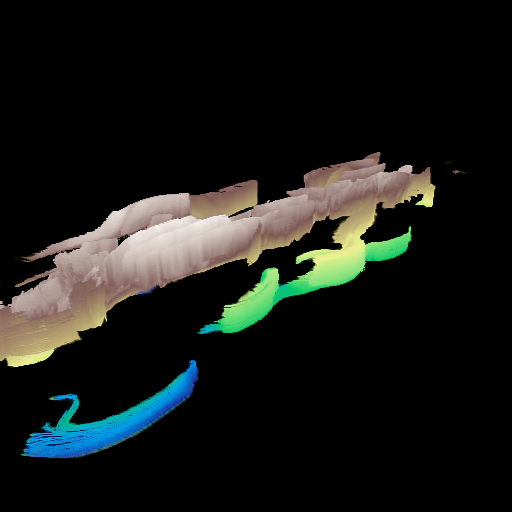} & \includegraphics[width=.17\linewidth,valign=c]{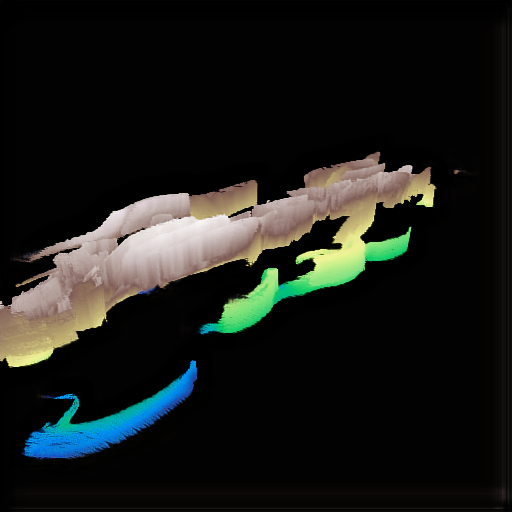} & \includegraphics[width=.17\linewidth,valign=c]{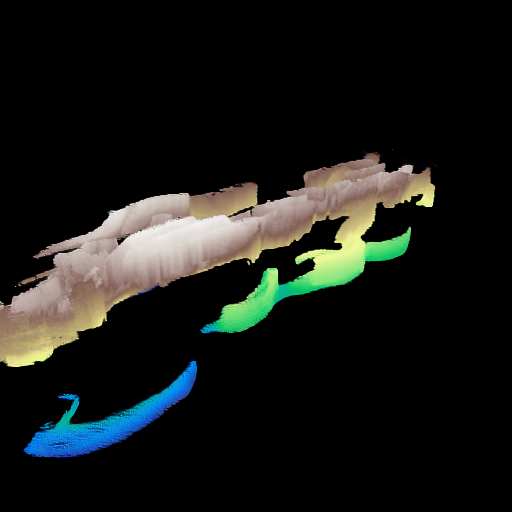}\\
    \vspace{3mm} 
    \rotatebox[origin=c]{90}{Storms} & \includegraphics[width=.17\linewidth,valign=c]{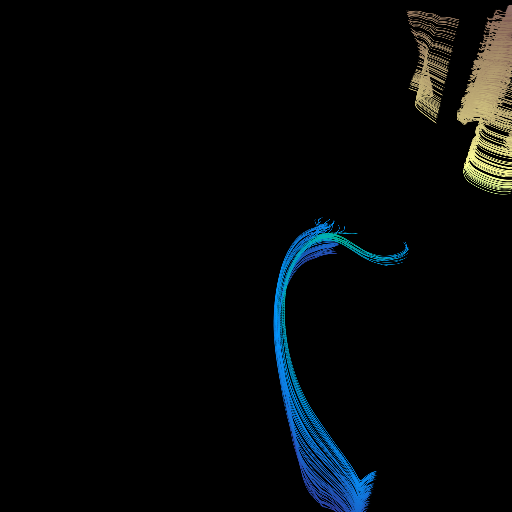} & \includegraphics[width=.17\linewidth,valign=c]{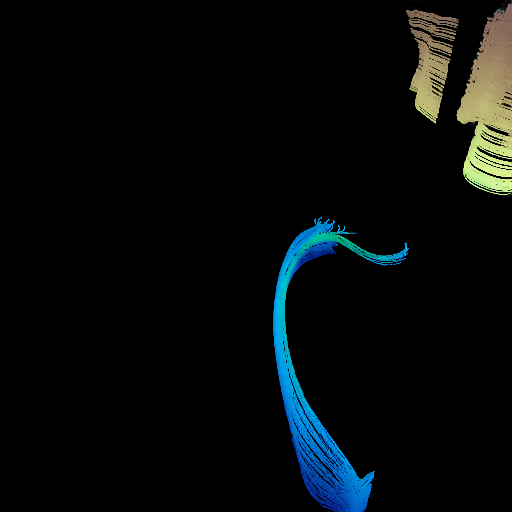} & \includegraphics[width=.17\linewidth,valign=c]{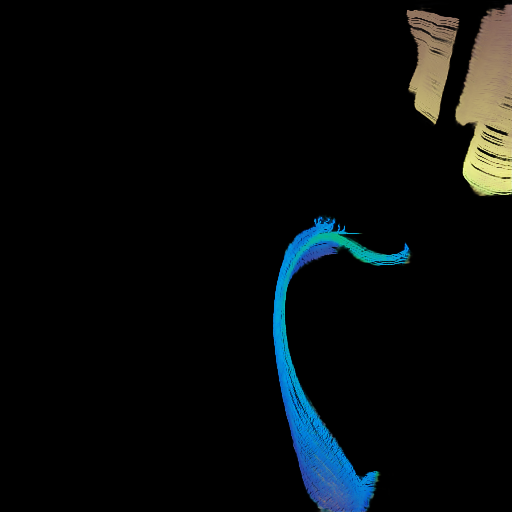} & \includegraphics[width=.17\linewidth,valign=c]{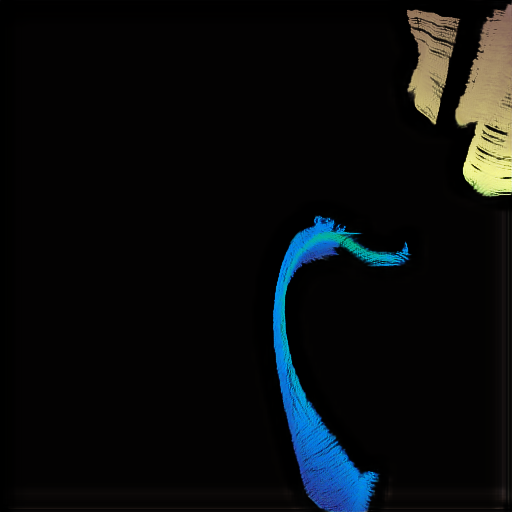} & \includegraphics[width=.17\linewidth,valign=c]{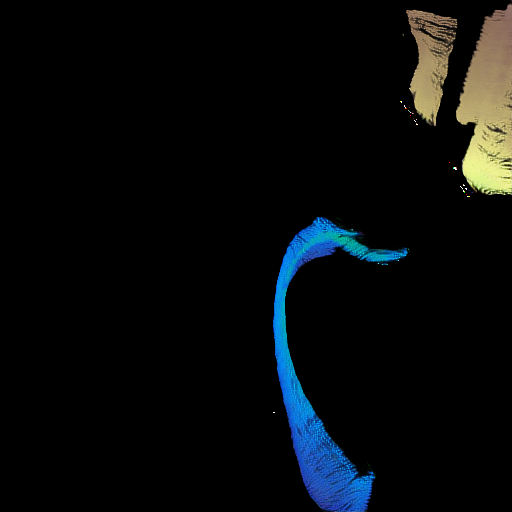}\\
    \vspace{3mm} 
    \rotatebox[origin=c]{90}{Morro Bay} & \includegraphics[width=.17\linewidth,valign=c]{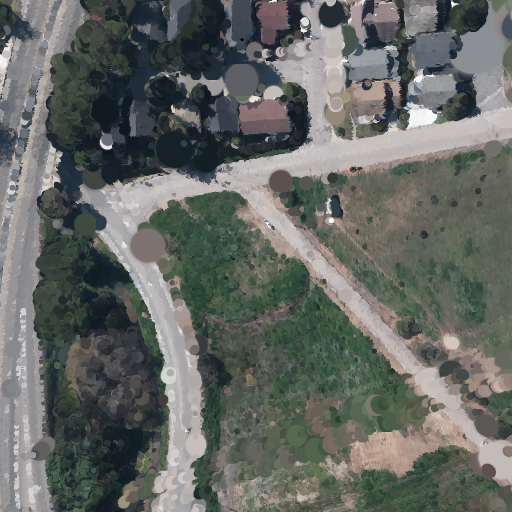} & \includegraphics[width=.17\linewidth,valign=c]{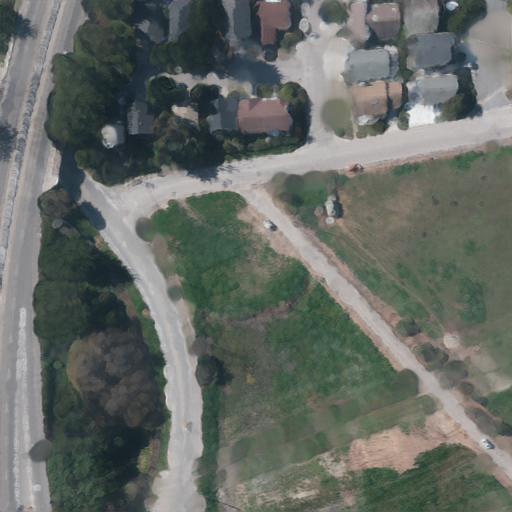} & \includegraphics[width=.17\linewidth,valign=c]{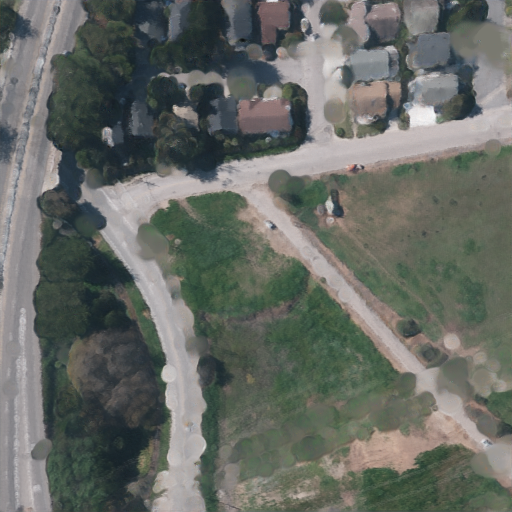} & \includegraphics[width=.17\linewidth,valign=c]{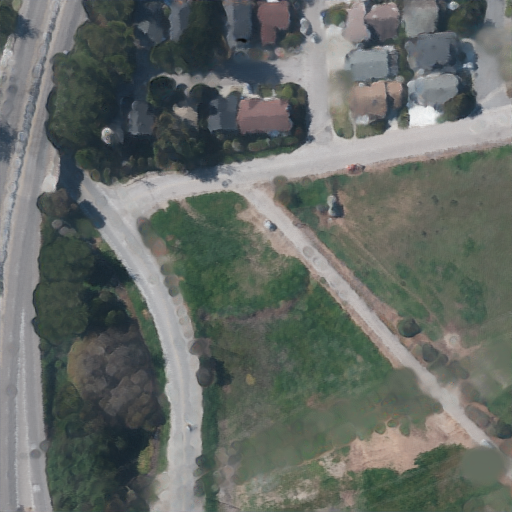} & \includegraphics[width=.17\linewidth,valign=c]{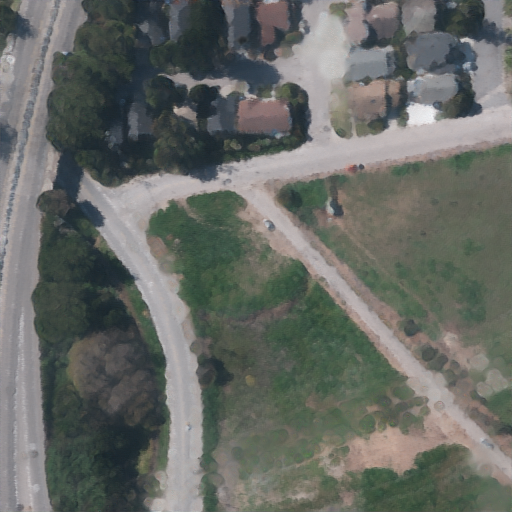}\\
    \vspace{3mm} 
    \rotatebox[origin=c]{90}{Morro Bay} & \includegraphics[width=.17\linewidth,valign=c]{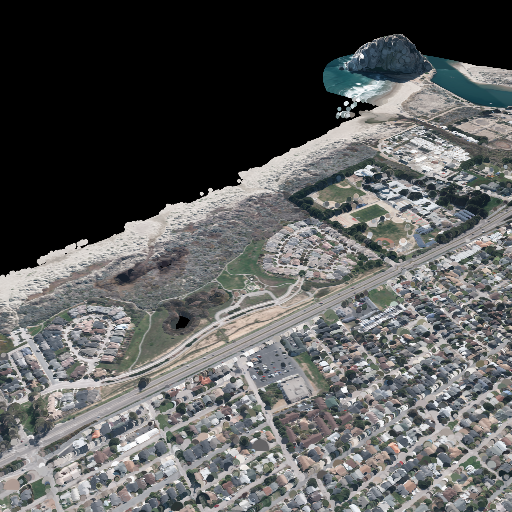} & \includegraphics[width=.17\linewidth,valign=c]{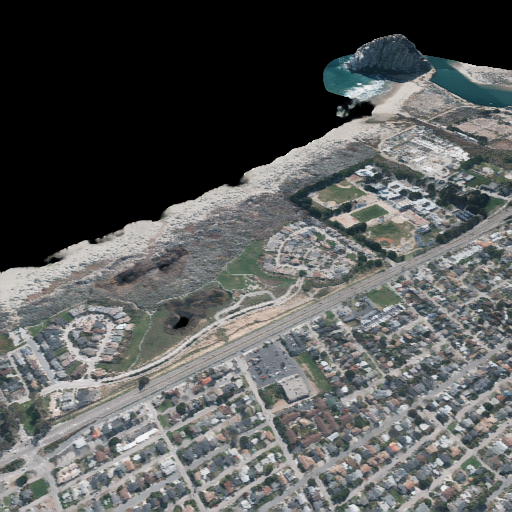} & \includegraphics[width=.17\linewidth,valign=c]{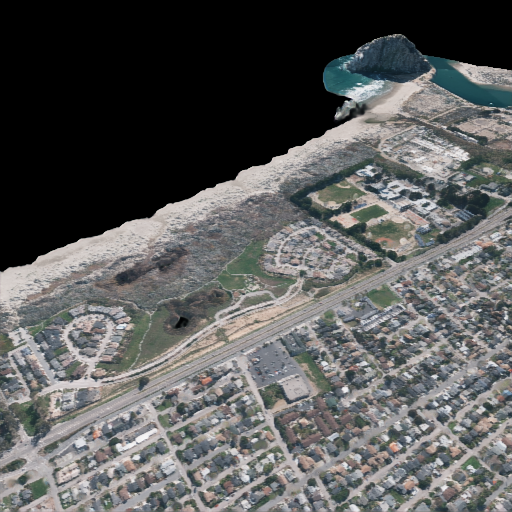} & \includegraphics[width=.17\linewidth,valign=c]{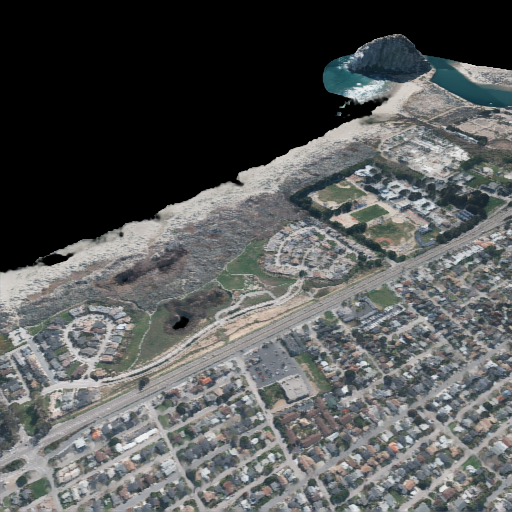} & \includegraphics[width=.17\linewidth,valign=c]{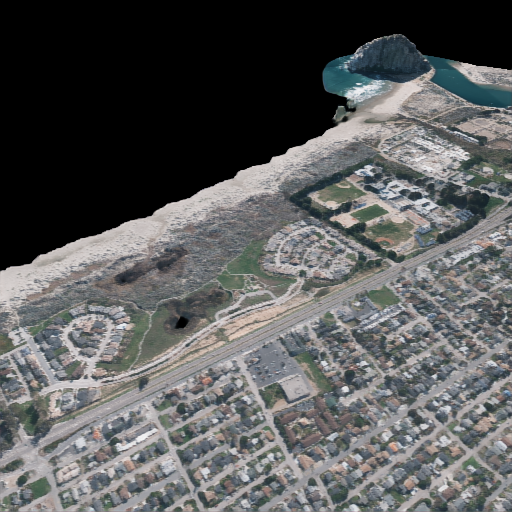}\\
    & (a) GSplat & (b)1x & (c) 2x & (d) 4x & (e) 10x
    \end{tabular}
    \caption{Qualitative results of NARVis at various PC sparsity when compared with the GSplat ground truth.}
    \label{fig:app_sr}
\end{figure*}

